\documentclass[twocolumn,showpacs,preprintnumbers,amsmath,amssymb]{revtex4}

\usepackage{graphicx}
\usepackage{dcolumn}
\usepackage{bm}

\begin{document}


\title{Modeling rf breakdown arcs}
\author{Z. Insepov, J. Norem$^*$}
\affiliation{Argonne National Laboratory, Argonne, IL 60439, USA}
 \email{norem@anl.gov} 
 
\author{A. Moretti }
\affiliation{Fermi National Accelerator Laboratory, Batavia, IL 60615, USA}

\author{D. Huang}
\affiliation{Illinois Institute of Technology, Chicago, IL 60616 USA}

\author{S. Mahalingam, S. Veitzer }
\affiliation{Tech-X Corp., Boulder, CO 80303, USA} 

\date{\today}

\begin{abstract}
We describe  breakdown in 805 MHz rf accelerator cavities in terms of a number of self-consistent mechanisms.    We divide the breakdown process into three stages: (1) we model surface failure using molecular dynamics of fracture caused by electrostatic tensile stress, (2)  we model the ionization of neutrals responsible for plasma initiation and plasma growth using a particle-in-cell code, and (3) we model surface damage by  assuming a process similar to unipolar arcing.   We find that the cold, dense plasma in contact with the surface produces very small Debye lengths and very high electric fields over a large area, consistent with unipolar arc behavior, although unipolar arcs are strictly defined with equipotential boundaries. These high fields produce strong erosion mechanisms, primarily self-sputtering, compatible with the crater formation that we see.  We use the OOPIC model to estimate very high surface electric fields in the dense plasma and measure these fields using electrohydrodynamic arguments to relate the dimensions of surface damage with the applied electric field.  We also present a geometrical explanation of the large enhancement factors of field emitters.This is consistent with the apparent absence of whiskers on surfaces exposed to high fields.  The enhancement factors we derive, when combined with Fowler-Nordheim analysis, produce a consistent picture of breakdown and field emission from surfaces at local fields of 7--10 GV/m.  We show that the plasma growth rates we obtain from OOPIC are consistent with growth rates of the cavity shorting currents using x-ray measurements.  We believe the general picture presented here for rf breakdown arcs should be directly applicable to a larger class of vacuum arcs.  Results from the plasma simulation are included as a guide to experimental verification of this model. 

\end{abstract}

\maketitle

\section{Introduction}   

Vacuum breakdown is one of the primary limitations in the design and construction of high-energy accelerators operating with warm (copper) accelerating structures such as muon colliders, neutrino factories, or the CLIC linear collider design  \cite{muprogram,mucool,MICE,CLIC}.  Nevertheless, there has not been agreement on the physics and the mechanisms that cause this phenomenon \cite{laurant}.  
  
Vacuum breakdown has a long history.  Starting with experiments done over 100 years ago by Earhart, Hobbs, Michelson and Millikan that first defined the process (see Fig. 1), and initial modeling by Lord Kelvin, to the present day, an enormous number of papers have been published, exploring all the experimentally accessible variables \cite{wood,earhart,lordk,hobbs,millikan,PR1,CERN1,CERN2,CERN3,CERN4}.  Nevertheless, there remains a lingering uncertainty about both the overall process and many of the experimental details, although there seems to be a similarity between arcs in different environments \cite{laurant,werner,alpert,schwirzke,Mesiats,Wilson,Latham,Dolgashev,andersbook,juttner1,juttner2,rohrbach,safa,jimenez}.

To a large extent, this uncertainty is due to the fact that events occur very rapidly, during which experimental parameters vary over many orders of magnitude and a large variety of mechanisms seem to be involved.  However, we also believe that, over the years, many assumptions have been made that are not generally true, and this situation has caused some confusion.  Among the behaviors that need  explanation is how these structures can operate for very long periods {\it without} breaking down.

\begin{figure}  
\includegraphics[scale=0.5]{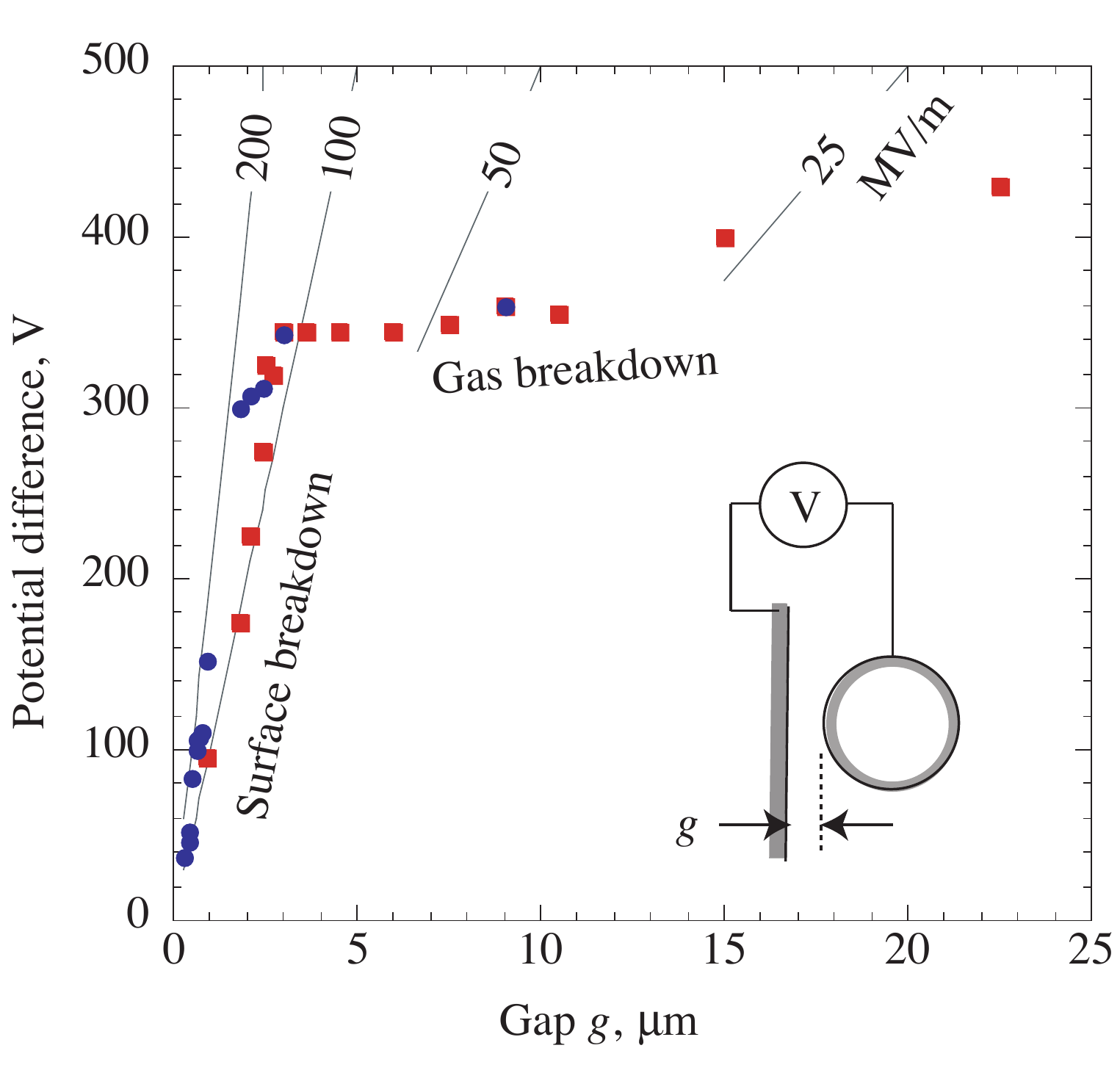}
\caption{Data from \cite{hobbs} and 1905 \cite{earhart}, showing that surface breakdown occurs at fixed electric fields when there is insufficient space for a gas breakdown avalanche.  This data, combined with the survey by Alpert \cite{alpert}, shows that vacuum breakdown is a different phenomenon from gas breakdown and a single surface phenomenon. }

\end{figure}
  
The assumptions  and conclusions presented in this paper are derived from experience with 805 MHz rf systems developed to explore muon cooling.  The MuCool test area at Fermilab has operated an experimental program to look at gradient limits in copper cavities since 2001 \cite{PR1,PR2,hassanein,palmer}.  This program has studied breakdown in open and closed cell structures, and the data obtained is our primary source of experimental input.  

This paper describes a model where breakdown is triggered by mechanical failure of a surface asperity, causing a charged metal plasma to form.  This arc can further heat the surface, producing more intense plasma and ionizing currents in an avalanche mechanism.  The plasma produces sufficient electron currents to short the cavity, discharging essentially all the cavity energy into the wall.  Both the electron beams and the arc, whose dimensions seem to be a few micrometers, can damage the surface, initiating subsequent breakdown events.  

We find that the arc must be described by a number of mechanisms.   Since the parameters of each mechanism are not well constrained by the data, we try to produce a complete model that we expect to be more tightly constrained and consistent than the individual mechanisms themselves \cite{PAC09,PAC07}.  We use molecular dynamics and the particle-in-cell codes OOPIC Pro and VORPAL to simulate the breakdown stages of the discharge, and we use analytical modeling to understand the properties of the high-density plasma during wall erosion \cite{OOPIC, OOPIC1, VORPAL}.  We find that the very low-temperature, dense plasma produced in arc simulations  produces conditions similar to those of unipolar arcs, with surface electric fields almost large enough to produce field emission and intense surface heating over comparatively large areas.  We use molecular dynamics  codes to simulate the high-density, low-temperature mechanisms in the breakdown arcs \cite{insepov}.  

Although our direct experience is mostly with 805 MHz systems, we  compare results from this modeling with experimental measurements of arcing in a variety of rf systems, small- and large-gap breakdown experiments, laser ablation, and tokamak results, where the mechanisms also seem applicable \cite{schwirzke,andersbook}.  The starting point for our work, shown in Fig. 2, is an example breakdown event in a 805 MHz pillbox cavity, as seen by x-ray monitors.  Since the x-ray flux is proportional to the rate of energy loss, $dU/dt$, in the cavity, we can reconstruct  the cavity energy, $U$, as well as a consistent picture of the cavity field and shorting current.  We expect that arc parameters depend on the geometry, frequency, location of arc in the plasma, and other variables; however, this figure shows the basic outline.

\begin{figure}  
\includegraphics[scale=0.22]{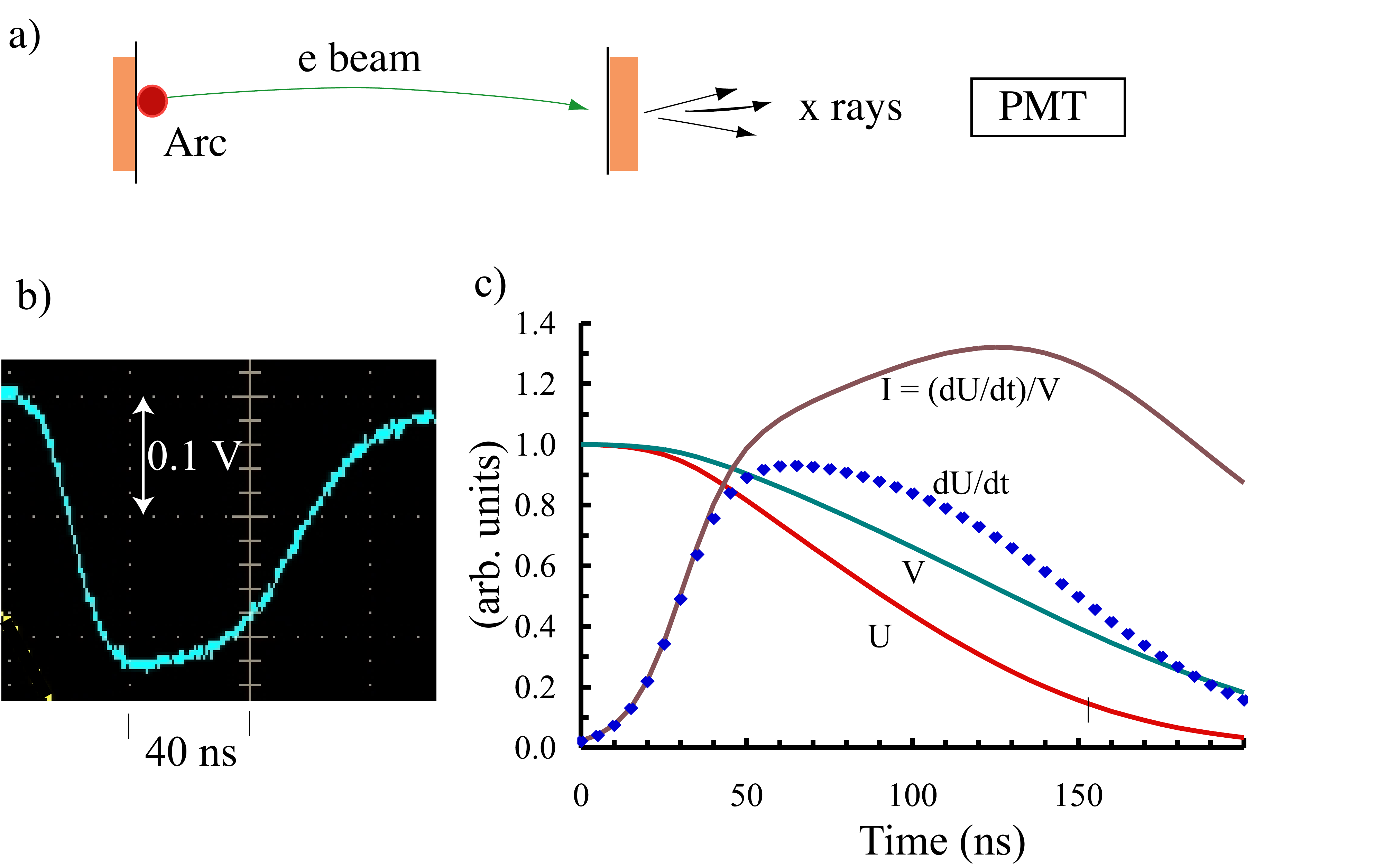}
\caption{ Breakdown in rf structures  (a) showing the arc (with dimensions of a few $\mu$m) that produces electrons that accelerate across the rf cavity, producing x-rays that are detected by the photomultiplier (PMT).  An example cavity breakdown event for an 0.805 GHz pillbox cavity can be reconstructed from x-ray data and the dimensions and parameters of the cavity (stored energy, $U_{max} \sim$ 1 J, potential difference, $V_{max} \sim$ 4 MV)  assuming the x-ray signal is proportional to the rate of energy loss.   From the PMT signal (b), we derive the approximate time evolution of the cavity parameters (c).  The figure shows the cavity stored energy, $U(t)$, the electron energies, $V(t)$, the shorting current, $I(t)=-(dU/dt)/V$, of order 4 A, and the radiated power, $P=-dU/dt,$ during a breakdown event, showing$\sim$ 10  MW EM energy leaving the structure. The PMT trace shows the signal with 40 ns/div and 100 mV/div, using a PMT supply voltage of 900 V.}
\end{figure}

This paper discusses the vacuum arc behavior that should be relevant to all geometries and applications.  Section II describes the surface environment and surface failure due to tensile stresses on asperities; Section III describes how the plasma is initiated, modeling the initial stages with particle-in-cell methods; and Section IV describes the evolution of the plasma to a very dense, cold state.  Section V describes the parameters of the non-Debye conditions we believe exist after the plasma evolves.  Little is known about these plasmas; however, we believe that experimental data on unipolar arcs is relevant both to the rf arc environment and to the interactions of dense plasmas with materials.  In Section VI we describe surface damage, considering self-sputtering and the possibility of field emission, which we believe can exist because of high surface fields.  Under some circumstances the surface field can be indirectly measured from the surface morphology of arc pits, and we find general agreement between the particle-in-cell predictions and measurements based on electrohydrodynamics.  We also find that sudden heating and cooling of the surface can produce cracks, which can produce many identical sites with field enhancements of the order seen in experiments.  We show that these field enhancements have very small heated volumes and very large thermal reservoirs so they do not heat up even with significant field emission current densities.  In Section VII we consider measurements of rise times of x-rays (shorting currents) and compare these rise times with results from the particle-in-cell simulation in Sections III and IV.  In Section VIII
we briefly compare other models, and
in Section IX we summarize our conclusions.
  
\section{The surface environment}   

X-ray measurements of field emission from a wide variety of rf cavities and coupons near breakdown limits show the existence of field emitters with surface fields of 7--10 GV/m \cite{PR1}.  Both the intensity and spatial distribution of these emitters have been measured at high fields from which the parameters and the distributions of the emitters can be extracted.  Measurements of field emitters in rf cavities and small spark gaps show similar properties.  There is also considerable experimental data on the distribution of the surface asperities that cause field emission obtained both from working rf systems and from extensive measurements with field emission microscopes.  Many examples are included in the literature; see Fig. 3 \cite{PR1,FN, nilsson,laurant,padamsee,mueller,PAC07,safa,jimenez}. 

Although the different types of data are not normalized to each other, they are roughly consistent with a parameterization of the form  $f(\beta)=$exp $(-D\beta)$, where $\beta$ is the enhancement factor, defined in terms of the local field at the asperity divided by the average surface field, and $\beta=E_{local}/E_{av}$, and $D$ is a numerical constant, approximately 0.03 \cite{PR2, nilsson,mueller,padamsee,mueller2}.  The data obtained by Nilsson et al. was the result of a technique designed to produce a distribution of high enhancement factors \cite{nilsson}.

Our understanding of the nature of field emitters comes primarily through measurements of dark currents and x-rays, using the model of field emission due to Fowler and Nordheim \cite{FN}.  This model, which explains the emission of electrons at high electric fields, was evidently the first precision prediction of quantum mechanics to be experimentally verified.  An earlier paper shows that the Fowler-Nordheim model correctly predicts the dependence of emission currents on applied electric field over 14 orders of magnitude, with the high currents being measured with a beam transformer and the low currents measured by counting individual electrons in a scintillator \cite{PR1}.  Over a narrow range of electric field it is possible to represent the Fowler-Nordheim emission curve with an expression $I_e \sim E_{local}^n$, where $I_e$ and $E_{local}$ are the emission current and local electric field, respectively.  Since the exponent $n$ depends on $E_{local}$, it was possible to directly measure $E_{local}$ in an operating rf system from a measurement of $n$, and we find $n\sim14$ for a wide range of experiments \cite{PR1}.

Since the experiments of Dyke et al. in the early 1950s \cite{dyke,barbour},  it has been frequently assumed that breakdown events were triggered by ohmic heating of sharp (at the 10 nm level) metallic whiskers on the conductor surface.  Whiskers have not been reported in either superconducting or normal surfaces in rf structures. However,  a high density of craters and other surface defects has been seen \cite{laurant,mueller}.  

Field emission, which requires local fields of 5--10 GV/m to produce measurable currents, frequently occurs in rf systems at average surface fields of 7--10 MV/m, requiring field enhancements on the order of 100--1,000, which are difficult to describe with some models \cite{juttner2,rohrbach}.  The most useful picture of emitter geometries and their relation to enhancement factors is obtained in superconducting rf studies and summarized in the book by Padamsee, Knobloch, and Hays \cite{padamsee}.  According to this reference, the experiments were done at Saclay with particles intentionally introduced into rf structures, where it was found that smooth particles did not field emit but rough-surfaced contaminants did \cite{safa,jimenez}.  The simple interpretation of enhancement factors proposed by the Saclay group, and others, is that the overall enhancement factor is a function of the product of the enhancement factors for the overall shape of the contaminant particle, $\beta_p \sim 10$, and the enhancement factor for micro-protrusions or microtips on the surface,  $\beta_m \sim 10$, so the overall enhancement is $\beta \sim \beta_p \beta_m \sim 100$, and the general mechanism is sufficient to explain the observed enhancement factors.  This is referred to as the ``tip-on-tip" model.  

Static electric field calculations have been done that support the idea that the field enhancement factors can be cascaded in this way \cite{noer}.  All the field emitting sites described in Ref. \cite{padamsee} show the jagged features at small scale, consistent with both significant local field enhancements at microtips and enhancements due to overall particle shape.  We find that structures at 805 MHz are very rough at the micron level, while our estimates of the size of field emitting surfaces have been been in the range of a few nanometers \cite{PR1}.

A general and straightforward explanation for the local enhancements of the electric field has been given by Feynman et al., as part of an explanation for gas breakdown that is also relevant to vacuum breakdown \cite{feynman}.  This argument derives field enhancements inversely related to the local radius, from Maxwell's equations.  We assume that the small radii are a result of statistical fluctuations due to particle contamination, surface cracking, and surface fracture.  Our own picture of field enhancements is given in Section IV-F.

The environment at a prebreakdown site is extreme in a number of parameters \cite{moretti}.  Measurements show that the local electric fields at asperities are in the range of 7--10 GV/m, leading to field emission \cite{FN,brodiespindt} current densities in the range of $10^{10}$--$10^{12}$ A/m$^2$. Tensile stresses exerted by the electric fields are in the range of 200--400 MPa.  These stresses are produced at a frequency of $2f_{rf}$, since the stress is proportional to $E^2$. In this environment it is not unexpected that materials will fail.   Imaging field emitters in the high electric field regions of well-conditioned cavities have shown that the surfaces are densely covered with asperities, and the field emitters seem to have similar brightness and hence similar field enhancements \cite{PR1}.  

\begin{figure}  
\includegraphics[scale=0.5]{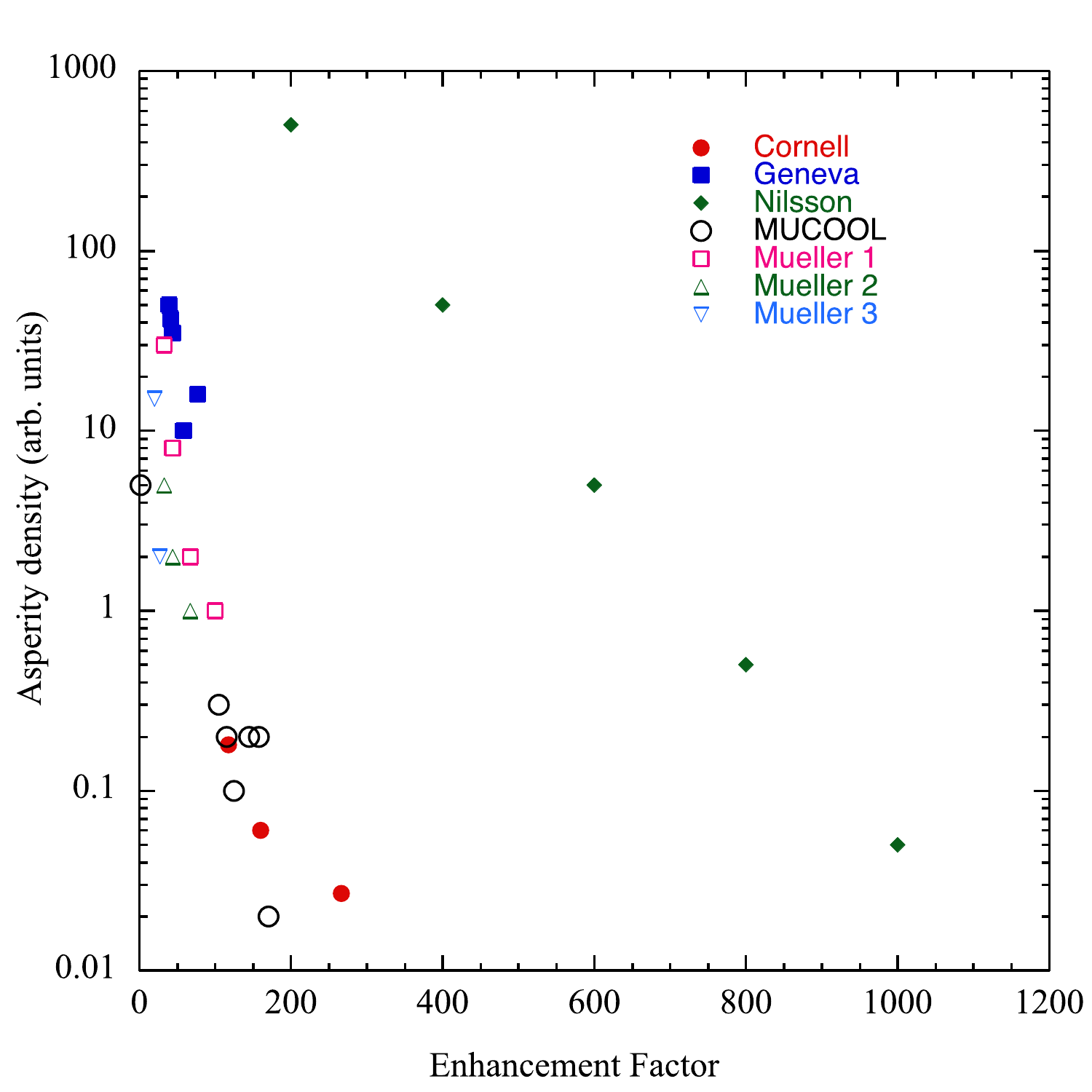}
\caption{Spectrum of enhancements in various ``clean" and ``damaged" surfaces.  This data is presented in the style of Nilsson \cite{nilsson} and includes data from Refs. \cite{PR2, nilsson,padamsee,mueller,mueller2,PAC07}.}
\end{figure}

\subsection*{Surface failure}   

We next describe a model of the geometry of the field emitters (breakdown sites) we expect are the cause of breakdown.  Our modeling shows that even with high densities of field emission currents, there is little heating of these emitters (see Section IV-F).  This result is consistent with imaging of field emitters that showed little change in the pattern or relative intensity of emitters over many weeks of operation of a cavity, implying that the emitters stay solid and do not deform \cite{moretti}. Thus we  assume that fracture, resulting from electrostatic tensile stresses, is the primary cause of failure, mainly because it is the simplest mechanism that can explain all the data.  Fracture in this context is relatively unexplored compared with the ohmic model that was carefully described 55 years ago in papers by Dyke et al.  \cite{dyke, barbour} and others.  The fracture mechanism is essentially identical to a Coulomb explosion of highly charged metallic clusters and molecules when electrostatic repulsion is greater than the binding forces, and this repulsion can cause them to burst apart, producing charged and very energetic ions, fragments, x-rays, and high-energy (keV) electrons \cite{belkacem,book}.  In the rf case, however, we expect fatigue (creep, at the atomic scale) to contribute because of the cyclic nature of the stress. This process occurs at surface electric fields of around 10 GV/m.  Coulomb explosions can occur in at least two contexts: (1) the initial fracture of the surface itself and (2) the breakup of fragments produced by fracture of surfaces in the presence of electron currents, which is the same phenomenon.  A practical advantage of assuming this trigger mechanism is that the process is basically one dimensional and can be more easily modeled.   This process has been described by using molecular dynamics \cite{insepov}.

Fatigue should be involved in the failure mechanism because asperities can operate for many cycles with stable operation before failing.  The relationship between failure rate and electric field is proportional to a high power of the electric field, which follows the laws of fatigue failure due to the pulsed tensile stress applied by the oscillating electric field.  The usual way to consider this problem is to look at the mean time between failures (MTBF) for components under cyclic stress.  The standard approach is to use the W\"{o}hler curve, which describes how the applied stress on a component affects the lifetime of that component \cite{wohler}.  Tensile stress-driven fracture at these fields has been modeled by Insepov, Norem, and Hassanein,  using molecular dynamics, showing how the electric field draws atoms carrying excess induced charges out of the material \cite{insepov}. 

Creep, which draws lattice defects toward areas of higher stress, creates cumulative damage in the material by creating points where the material can yield.  Measurements at CERN of MTBF for small gap systems are consistent with fatigue predictions \cite{CERNnorem}.

While ohmic heating may occur, and Dyke et al. and Trolan et al. have argued that it is likely in more resistive tungsten needles, the geometry matters because the volume in which ohmic heating occurs decreases very rapidly as the cone angle of the field emitter increases, and the thermal mass to be heated increases with the cone angle \cite{dyke,barbour}. 

Fracture followed by ionization of neutrals can be modeled in a straightforward way and explains the initial stages of the arc. Thus we are not concerned in this paper with alternative mechanisms.  Atom probe tomography, which studies materials at very high positive surface potentials, sees materials fracture but does not see significant motion of surface atoms or gas production in 1--50 M pulses at $\sim$ 10 GV/m field levels  \cite{hassanein,miller}.  Ultimately, however, all the mechanisms outlined above are caused by the same local surface fields and are indistinguishable after the arc begins.   While melting is not required in the initial breakdown trigger, we expect melting in the subsequent development of the arc.   

\section{Plasma Initiation}    

We assume that the trigger for breakdown events is the injection of high-density material above a field emitter, where the intense field emission currents would break up and ionize the material to produce a plasma.  A number of mechanisms could be involved.  Coulomb explosions  could result from either ionization or electron absorption of field-emitted electrons by clusters.  Lord Rayleigh showed that when the electrostatic energy becomes comparable with the binding energy of the cluster, it can become unstable \cite{rayleigh}.   The number of charges involved in Coulomb explosions of small (10--100 nm) clusters is  on the order of 1000 electrons, a small fraction of the $\sim$6 M electrons/ns field emitted in a 1 mA  field emission current.  We expect that the Coulomb explosion process, which may involve multiple ionizations, would be sequential until the solid material is reduced to atoms or ions.

The dimensions of asperities have been indirectly measured and found to be approximately a few nanometers radius \cite{PR1}.   Thus the electron energy, $U$, for the field-emitted beams would be in the range $U  = e(E \sim 10  $ GV/m$)(dr \sim 5$  nm$) \sim$ 50\  eV.  Electrons in this energy have a range on the order of a few nm in copper.  We show in the following section that the electric field is locally enhanced by the plasma sheath in addition to the geometrical enhancement produced by the local geometry.  Thus, kinetic energy derived from the electric field would be efficiently deposited in small clusters. 

\subsection{Modeling with OOPIC Pro}

Ionization of neutral metallic gas has been modeled by OOPIC Pro assuming field-emitted electrons are produced below an inertially confined atomic gas \cite{OOPIC1,werner}.  Initial results show that the ionized electrons, as well as the majority of the field-emitted electrons, are accelerated through the plasma, producing a net positively charged plasma, which is slowly expanding because of its own charge.

OOOIC Pro is a particle-in-cell physics simulation code for 2D (x, y) and (r, z) geometries with 3D electrostatic and electromagnetic field solvers and Monte Carlo collision and ionization models. The code has been extensively benchmarked, operates on many platforms, and has a useful user interface \cite{OOPIC1}.  The results shown in this paper do not seem to be strongly configuration or parameter dependent.  To study the breakdown process in a simple way, we have modeled a geometry where a cylindrical cloud of neutral copper gas 2 $\mu$m thick is suspended inertially 1 $\mu$m over a field-emitting asperity, as shown in Fig. 4.  The overall length and radius of the volume considered are 10 $\mu$m, and the applied electric field has a frequency of 805 MHz.

The copper gas models a copper fragment broken off the tip of an asperity.  The dimensions of the conical asperity are 2 $\mu$m in height and 4 $\mu$m in diameter at the base in order to localize the plasma.  Such conical asperities, while not unphysical, are not a necessary component of the model,  but they are useful for computational purposes.  We assume that a field emitter is located on the surface of this asperity such that the enhancement factor of the combined system is 184 \cite{PR1}.  OOPIC Pro models field emission at high current densities in a self-consistent way by calculating space charge fields in the presence of plasma ions and electrons.  The ionization of copper and various secondary emission coefficients are contained in the code \cite{OOPIC}.  The grid size for these initial runs is set at 200 nm, and the time step is set at $10^{-14}$ s, which seems adequate for plasmas whose dimensions are a few nanometers. The processes considered by OOPIC Pro are shown in Fig. 5.

\begin{figure}   
\includegraphics[scale=0.55]{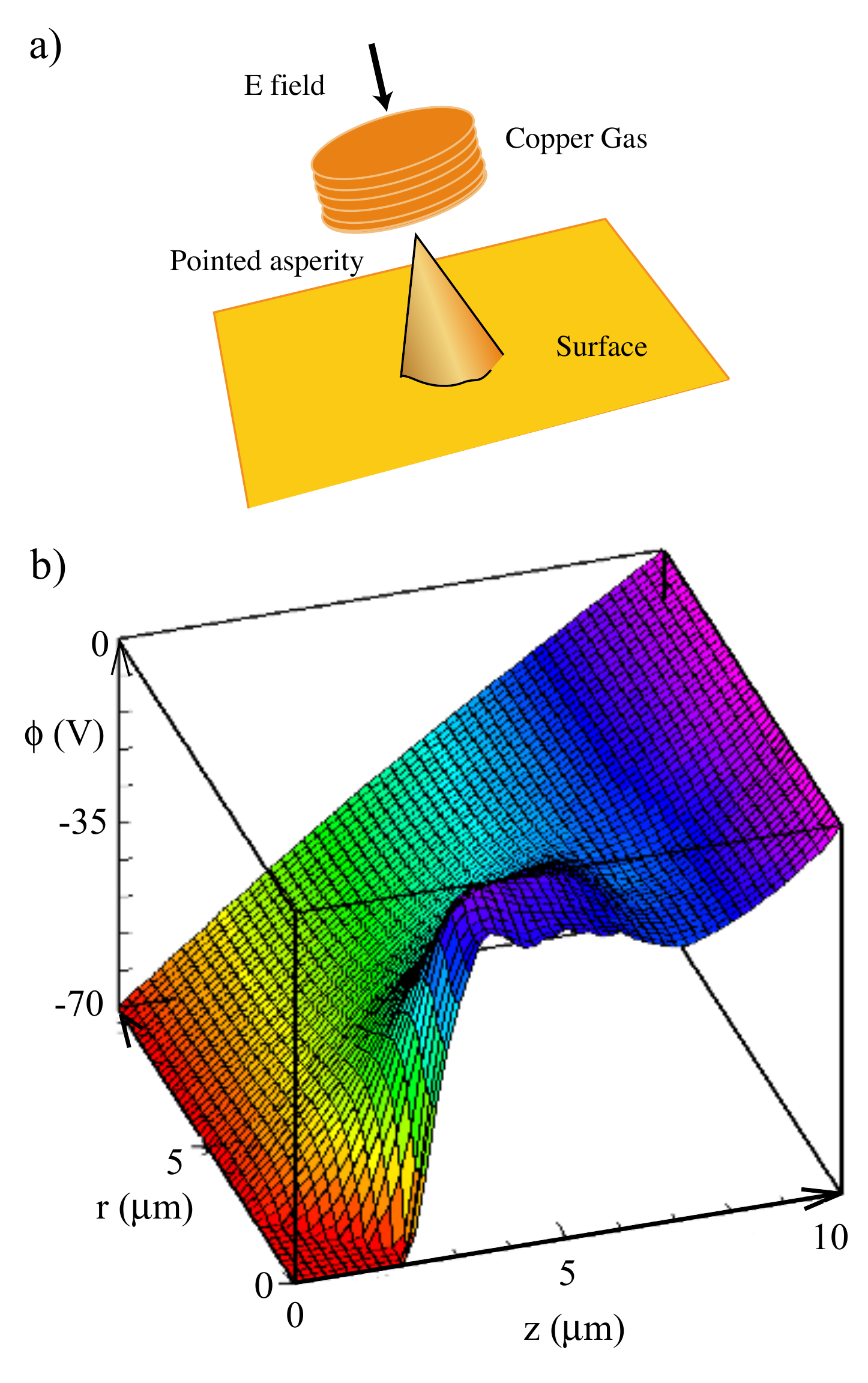}
 \caption{(a) Basic geometry used in these calculations:  an asperity field emits electrons into a copper gas; (b) OOPIC Pro results showing the scaler potential $\phi(r,z)$, in the region of an asperity located at radius $r=0$ and distance $z=0$.  The plot shows how the surface electric field, the slope of the potential,  $E_z=d \phi / d z$, is amplified by the positively charged plasma that forms over the asperity.  The effect of the plasma formation is to almost immediately increase the local electric field at the breakdown site by a large factor over the field, $E_{local} = \beta E_{surf} \sim $7 GV/m, with $\beta = 180$, which caused the initial breakdown trigger.}  
 \end{figure}         
	
 \begin{figure}   
 \includegraphics[scale=0.75]{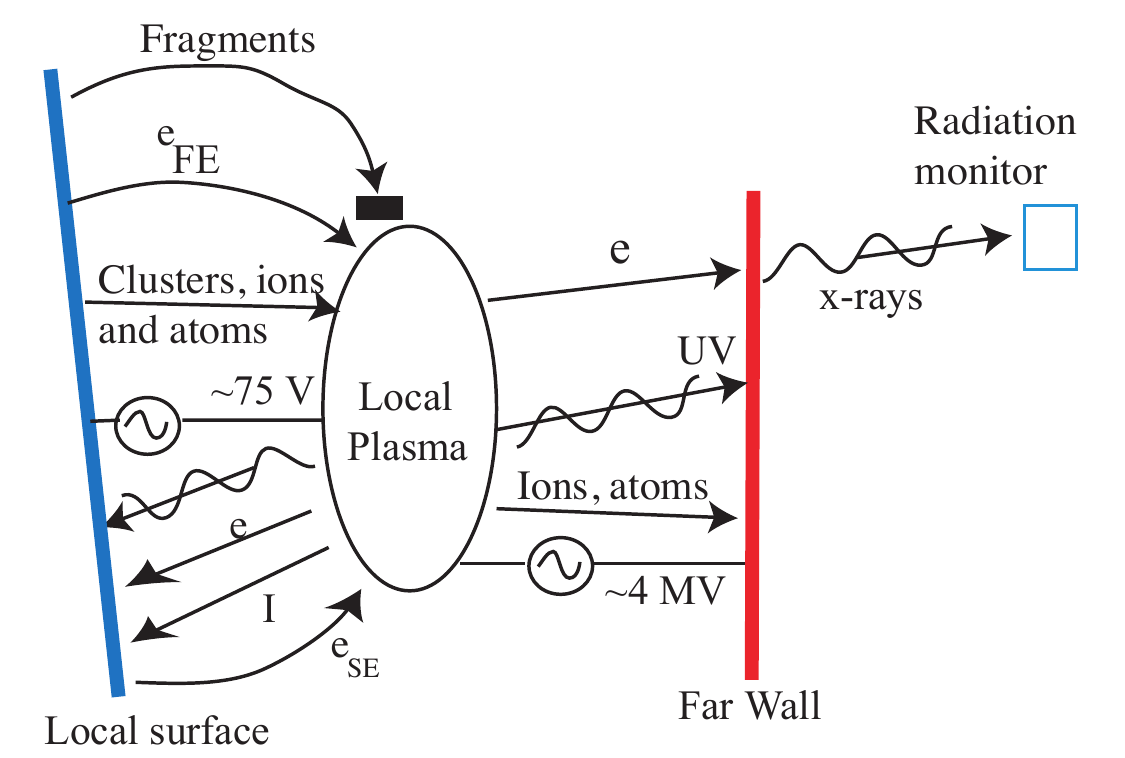}
 \caption{Energy and particle flows during breakdown.  We assume the process is initiated by the production of fragments that are heated and ionized by field-emitted electrons, producing a plasma whose electrons can be accelerated to the far wall to short the cavity fields.  The dimensions of the plasma are on the order of tens of $\mu$m, with an electron tail extending to the far wall.}
 \end{figure}
 
The initial evolution of the discharge is plotted in Fig. 6, which shows how the numbers of simulated ions and field emitted, trapped, and secondary electrons  evolve in the first 5 ns of the discharge, using the model described in Figs. 4 and 5.  The ion energy and ion density of the arc are shown in Figs. 7 and 8 at 7.3 ns into the discharge.  They show how the ion cloud builds up from cold, neutral gas.  Ions at the edge of the plasma are ballistically accelerated away from the arc toward the walls by the space potential of the plasma..
 
\begin{figure}   
\includegraphics[scale=0.3]{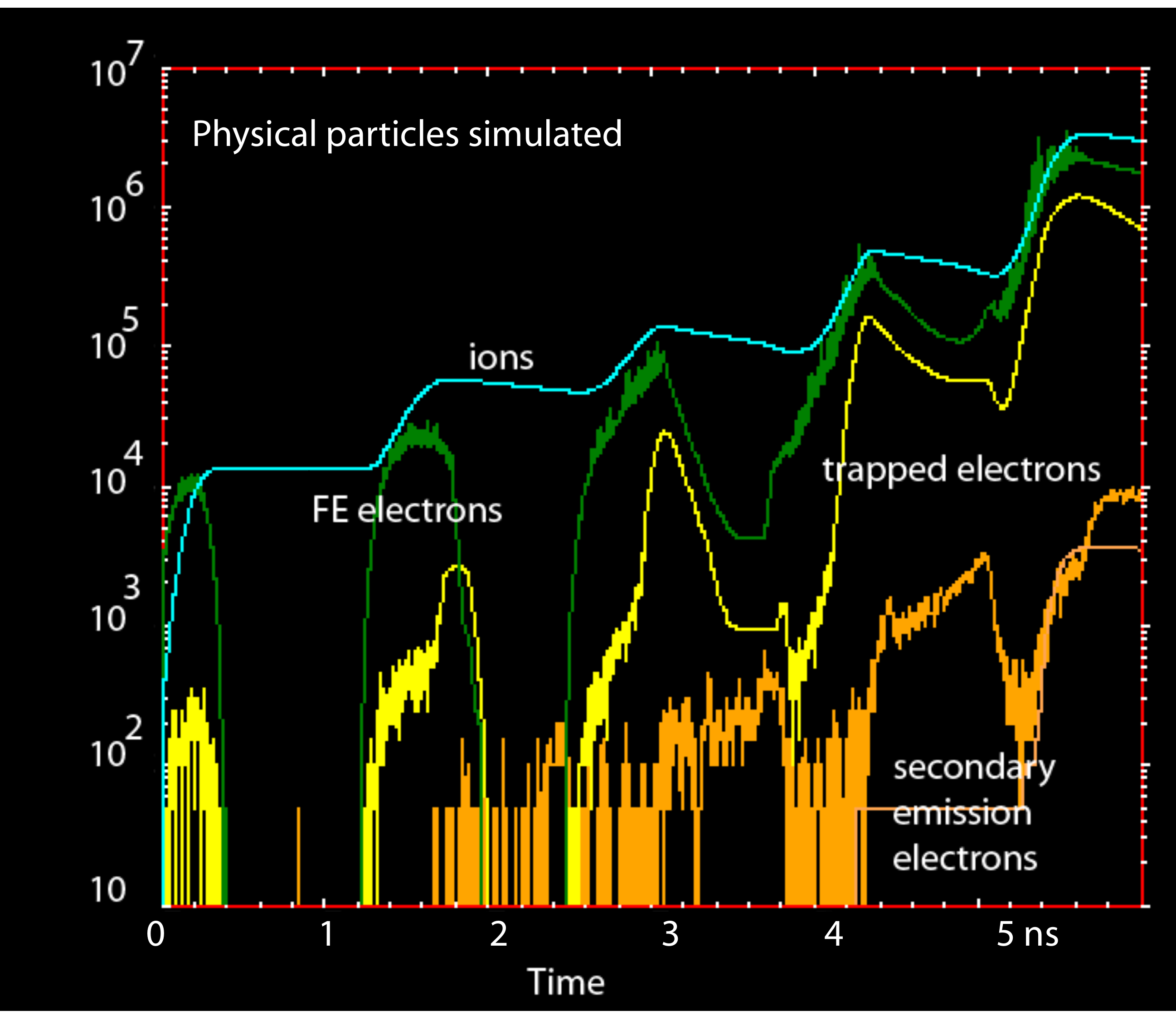}
 \caption{Plasma initiation near threshold.  Green is the number of simulated field-emitted electrons, blue is the number of ions, yellow is the number of plasma electrons, and orange is the number of secondary emission electrons.}
 \end{figure}

 \begin{figure}   
 \includegraphics[scale=0.44]{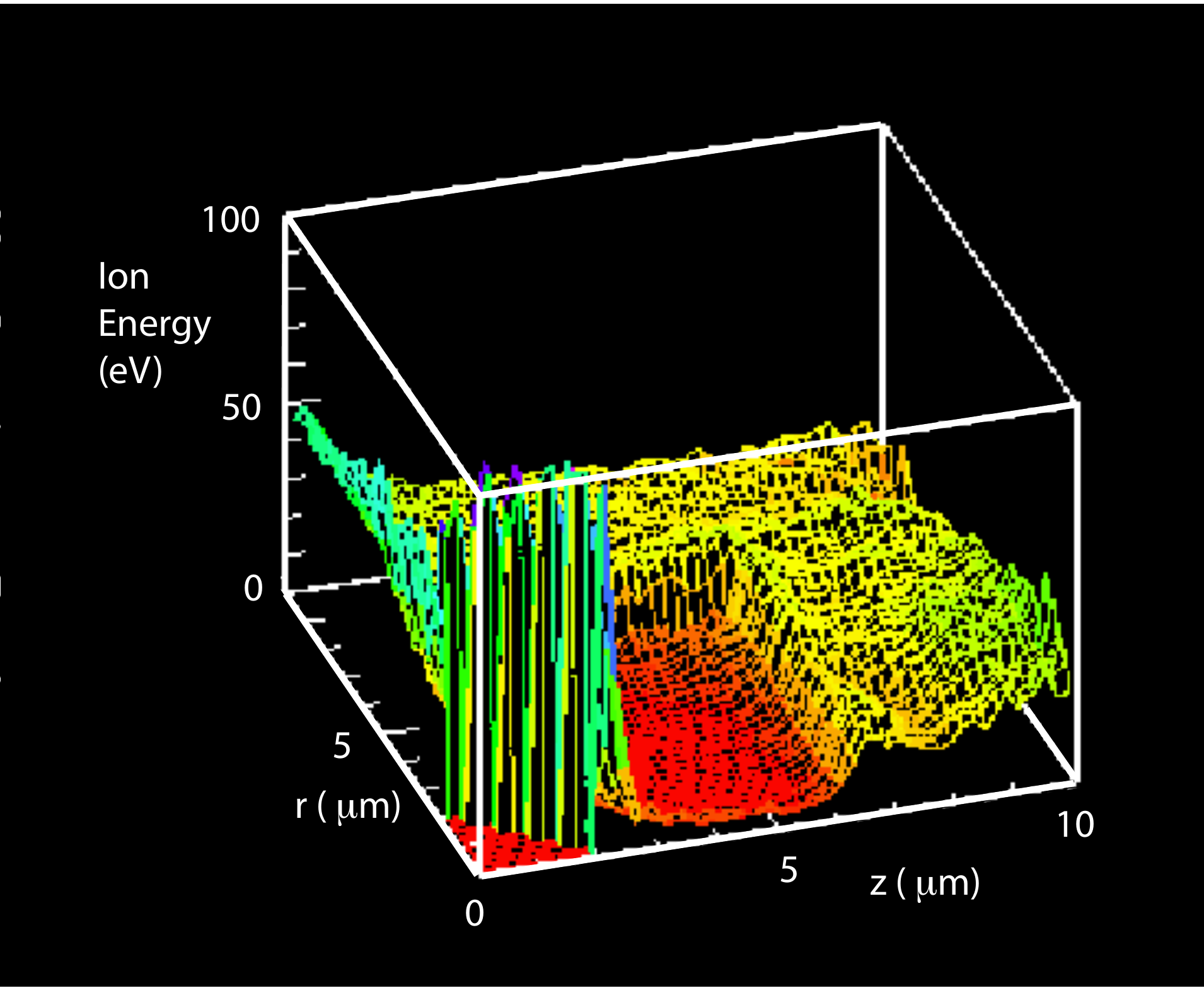}
\caption{Ion energy as a function of the potential $\phi(r,z)$.  Ions are created above the field emission source at very low temperature and  are accelerated outward in all directions by their self-field.  The plots show the ions streaming away from their source of ionization. }
 \end{figure}
 
 \begin{figure}   
 \includegraphics[scale=0.57]{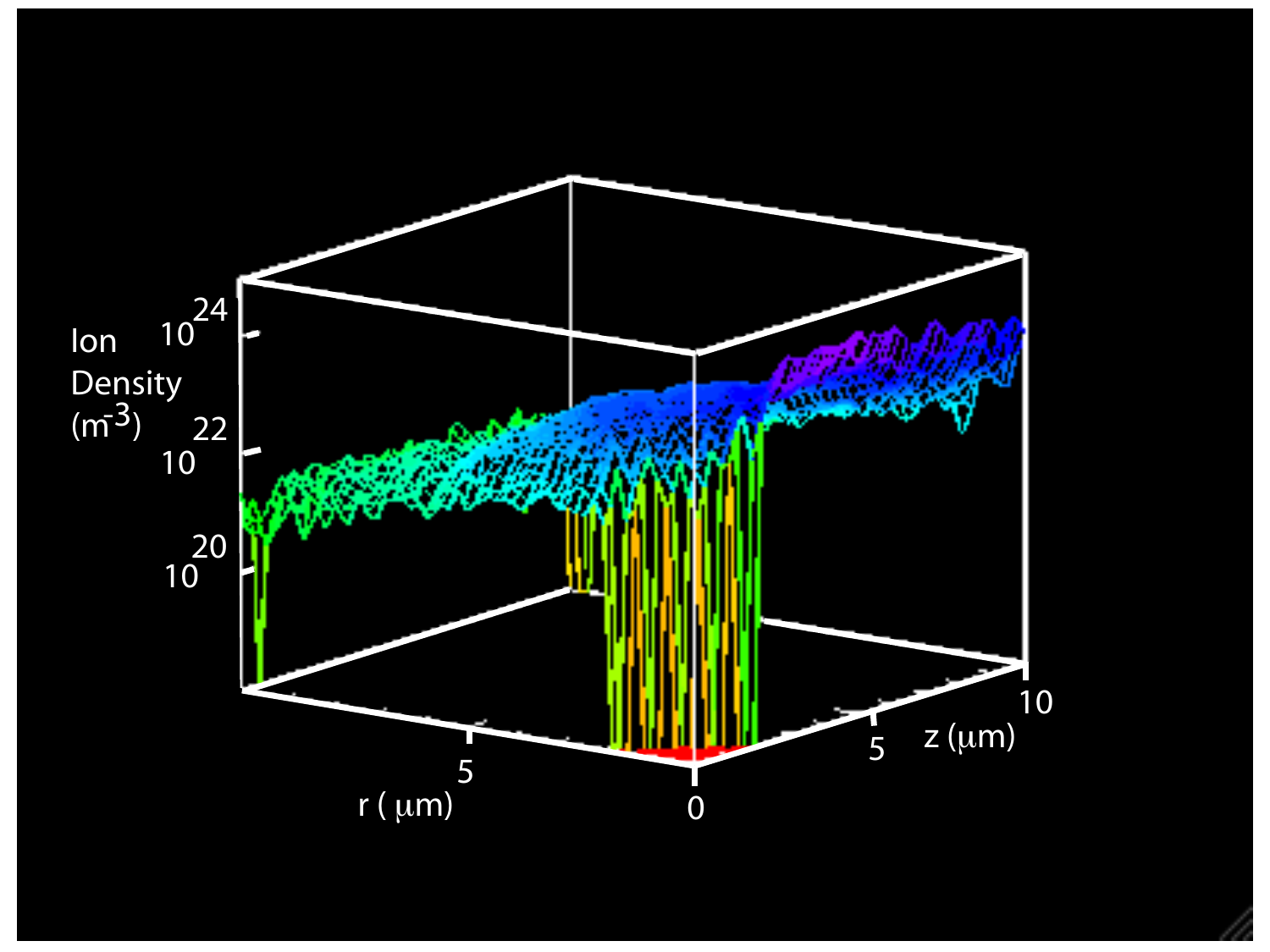}
\caption{Ion density in the arc.  The ions in the arc are inertially confined against the potential $\phi(r,z)$, which tends to blow up the ion cloud.  }
 \end{figure}

Since the OOPIC Pro code has been used in a number of different applications and has been benchmarked against other codes, the initial aim of this  preliminary effort was to calculate radiative losses due to line and continuum radiation, in order to understand the complete spectrum of heat flux on the wall.  Although the code cannot yet follow all the orders of magnitude of changes in density and current, it should be fully functional in its ability to provide a description of the first 7.5 ns of a discharge, giving the plasma dimensions, densities, electron and ion temperatures, and power fluxes to the wall due to ions and photons; see Fig. 6.   The code can describe the initial ionization of neutral gas and the exponential growth during the first six orders of magnitude of ion and electron behavior in the arc. The growth rates seen in the simulations are consistent with experimental data, as described below.  While the simulation is not sufficient to completely describe the whole arc behavior, we present a number of mechanisms for fueling and limiting the arc, which we believe would control the arc behavior as the arc develops further.   We assume that we can extrapolate further development of the arc, based on the behavior seen in the first 7.5 ns; and experimental data confirms this.

Simulations in the case of rf discharges show that the plasma arc has dimensions on the order of a few microns. As the discharge develops, the center of the arc reaches high densities, causing the plasma frequency 
 \[f_{[Hz]}= \sqrt{n \  e^2 / \epsilon_0\  m} = 3 \times 10^{13},\]
which is much higher than the rf frequency of the cavity itself, to screen the center of the arc plasma from the applied rf fields. 

 \subsection{Threshold neutral gas density}
 
 By varying the neutral density one can estimate the sensitivity of the ionization process to the pressure of the neutral metal gas.  While most of the simulations have been done at an initial pressure of 5000 pascals, we have looked at the time dependence of the breakdown process as this pressure is increased and decreased by a factor of approximately 3~\cite{PAC09}. Increasing the density breaks down the gas more rapidly, reducing the time constant for the density increase by more than  the ratio of the pressure increase.  Decreasing the pressure of the neutral gas by a factor of 3 produces a stable configuration, where there is almost no increase in the ionization.  Thus, the initial gas pressure has  a strongly nonlinear threshold effect on the initial time dependence of the ionization.
 
The neutral pressure of 5000 pascals, combined with a thickness of 3 $\mu$m, is equivalent to the mass in the top half monolayer of the copper.  This density also corresponds to that predicted by the argument that the ionization cross section times the density and path length, $\sigma_i n dz,$ should be on the order of 1 \cite{werner}.   It is interesting to compare the required mass of atoms with the one monolayer coverage of copper oxide, found on a copper sample that had been cleaned and stored in air for a week before being measured by atom probe tomography at high surface fields \cite{hassanein}.  It is also interesting to speculate whether the sudden removal of the tightly bound oxide layer could be responsible for breakdown of copper structures.

\subsection {Trapped electrons} 

Figure 9 shows the behavior of trapped electrons in the ion space charge field.  The plot shows the $v_z$ vs. $z$, with field-emitted electrons colored green, plasma electrons colored yellow, and ions colored blue.  The field-emitted electron velocities roughly follow the shape of the potential function shown in Fig. 4, and, on this plot, the trapped electrons rotate around the center of the space charge potential, with a period on the order of 10 ps.  The fluctuations in the field-emitted electrons are caused by operating near the space charge limit.  The trapped plasma electrons contribute significantly to the ionization of neutral gas.

\subsection{Plasma size}
The plasmas modeled by OOPIC Pro have radial dimensions on the order of a few nanometers and lengths, perpendicular to the surface (or somewhat larger, depending on the presence of  a magnetic field).  The contact area with the surface where plasma pressures, melting, and other damage mechanisms are significant is also on the order of a few nanometers.  These dimensions are similar to many measurements of damage sites for single breakdown events as measured by a number groups; see Fig. 3.28 in \cite{andersbook}.  For highly damaged systems where damage sites could be piled on top of each other, these sites could be much larger.  We show in Section IV that arcs can produce a wide variety of damage depending on how they are powered; thus, the damage site is not a reliable way to measure the arc dimensions except in special cases.

\begin{figure}  
\includegraphics[scale=0.54]{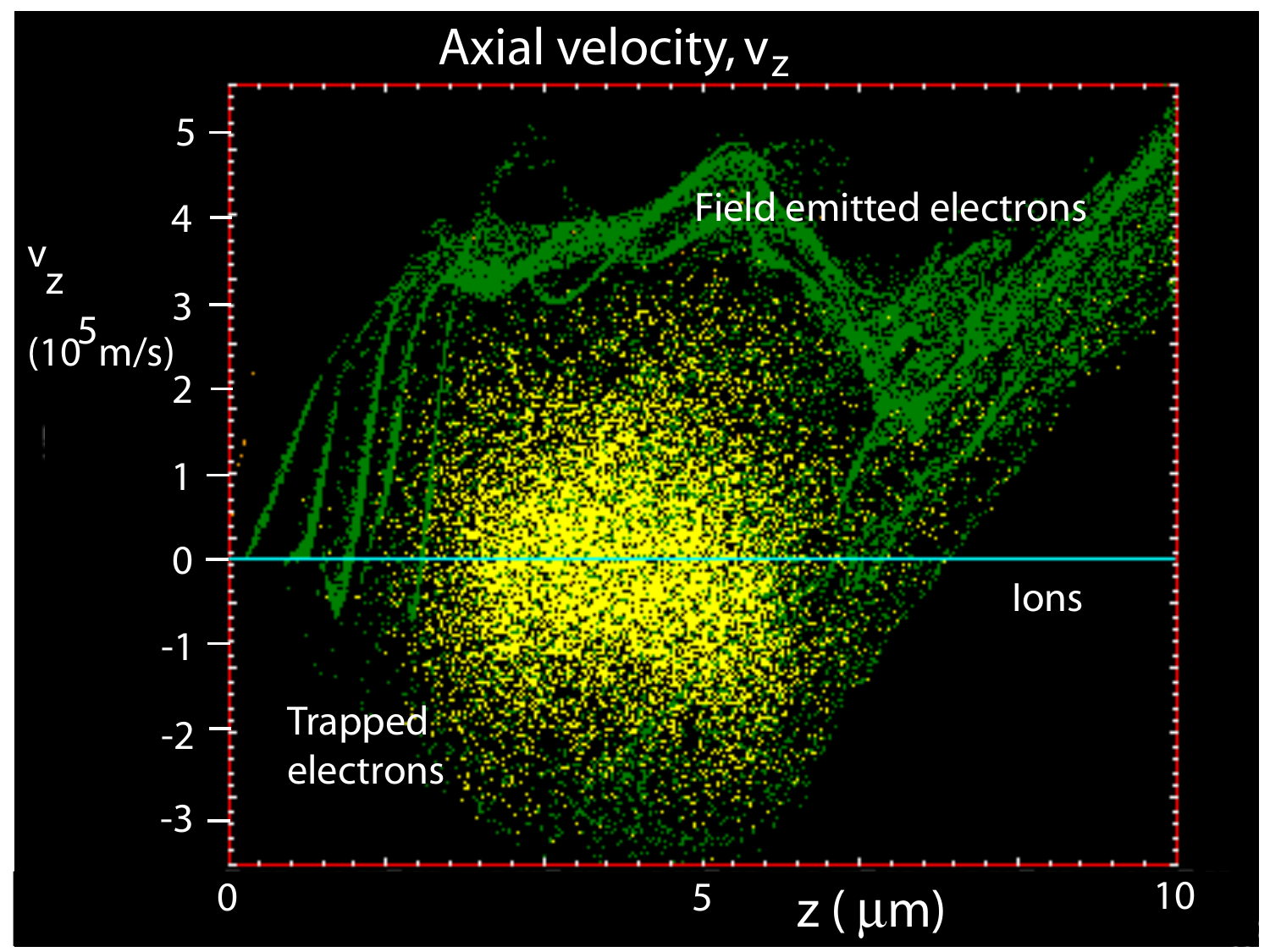}
\caption{Plot of velocity vs. distance, $v_z$ vs. $z$, for all species, with field-emitted electrons colored green and plasma electrons colored yellow and ions being blue. The trapped electrons circulate around the center of the potential with a period of about 10 ps.  These trapped electrons are an important source of ionization in the later stages of the arc. }
\end{figure}

\section{Plasma Evolution}    

In this section we describe the mechanisms that drive the evolution of the plasma, from the first ions produced until the density increases beyond our ability to model it.  Modeling with OOPIC Pro shows the basic mechanisms that operate in a metallic plasma in a high-gradient environment driven by field emission from a nearby source.   The growth of the plasma in the rf environment is essentially exponential, 
\[ n \sim e^{cN},\] 
where the particle densities, $n$, are a function of the number of rf cycles, $N$, and some constant, $c$, where the time constant varies somewhat as the discharge evolves.  Electrons are swept away from the ions by the background electric field, leaving a positively charged cloud of relatively slow-moving ions; see Fig. 4b.  The net effect is to increase the surface gradient seen at the asperity, significantly increasing the field emission current and tensile stresses.   

\subsection{Space potential and surface electric field}
 
The function of the plasma sheath in static laboratory or fusion plasmas is to form a potential barrier so that the electrons  are confined electrostatically.  This potential barrier adjusts itself so that the flux of electrons and ions leaving the plasma is equal to maintain neutrality \cite{chen}.  Thus, the normal plasma sheath for a static plasma depends primarily on the electron temperature.  

In these simulations, ions are produced in a vacuum by a field-emitted beam, and these ions are inertially confined to region where the ionization occurs.  The electric field of the cavity sweeps the electrons produced in ionization (as well as the field-emitted electrons) away from the near wall, producing a positive space charge potential.  As the density of this ion cloud increases, the potential increases until it is able to trap ionization (and field-emitted) electrons; but throughout the rf cycle, electrons are primarily lost to the far wall of the cavity, maintaining quasi neutrality.  Those electrons that do hit the near wall do so with very little energy and only at the wrong polarity for field emission.  Thus the sheath potential is an artifact of ionizing the gas near a surface,  producing a localized charge,.

As the arc evolves, simulations show that while the plasma density increases approximately exponentially, the electron and ion temperatures do not significantly increase. The Debye length, 
  \[\lambda_{D\  [m]} =\sqrt{\epsilon_0 k T_e/n\ e^2},\] 
decreases, where $\epsilon_0$, $k$, $T_e$, $n,$ and $e$ are the permittivity of free space, the Boltzmann constant, and the electron temperature, density, and charge, respectively.  As the density passes $10^{24}$ charges/m$^3$, $\lambda_D$ approaches 20  nm for electron temperatures of 10 eV.   From the electrostatic potential of  50--100 V produced by the OOPIC simulations, which is consistent with the burn voltage of copper arcs \cite{andersbook}, this results in a surface field 
  \[E_{[V/m]} \sim \phi/\lambda_D \sim 3 - 6 \times10^{9}, \] 
a range that is capable of inducing field emission, tensile failure of materials, ion bombardment, and other mechanisms of damage, even in cold surfaces---and these surfaces would be hot as a result of ion bombardment.   On the other hand, the plasma parameter, defined as the number of electrons in a Debye sphere,
decreases to small values, $\sim 1$, as the density increases. Thus, efficient screening is not always possible, and the problem must be considered by using numerical methods.  The relevant parameter then becomes the plasma thickness required by Gauss's law to contain the charge required to support a given surface field, $E$. This thickness is equal to 
\[d_G = \epsilon_0 E / e n.\]
For cold plasmas with densities on the order of $10^{25}$ m$^{-3}$ and  electric fields less than or equal to 1 GV/m, $d_G$ is  comparable to or smaller than $\lambda_D$. Thus, screening would be somewhat effective.  Under these conditions, a large area (square microns) would function like an active field emitter, subject to Fowler-Nordheim emission and thermal emission while being actively eroded by ion self-sputtering.  Modeling shows that electrons introduced into the arc would continue to ionize neutral atoms, increasing the flux of ions hitting the surface.  A significant fraction of field emitted electrons would traverse the sheath without scattering, heating the plasma \cite{schwirzke}.As the surface field increased above 1 GV/m, screening would be ineffective.

\subsection{Plasma radiation and surface heating}

As an example of the time development of the plasma, we show in Fig. 10 an OOPIC simulation of the flux of line radiation on the asperity in the early stages of a discharge. The ionization state is determined from coronal equilibrium, for a given electron energy and atomic species \cite{mosher}.  As the plasma develops with time, the densities of both the electrons and ions increase; and the line radiation, which is proportional to the product of the two densities, also increases.  The generation of a dense plasma would require a significant time before measurable currents could be generated and the breakdown event could be externally detected from the loss of cavity energy.  Figure 6 shows a roughly exponential growth of the plasma density (the square root of the radiation level) with a time constant on the order of 1 ns.     The ``impurity radiation" shown in Fig. 10 can come from two sources: line radiation, proportional to $n_in_eT_e^{-1/2}$, and continuum radiation, proportional to $n_in_eT_e^{1/2}$, where $n_i, \ n_e,$ and $T_e$ are the ion, electron densities, and electron temperature, respectively.  At the low temperatures predicted by the model,  line radiation dominates.   

\begin{figure}  
\includegraphics[scale=0.55]{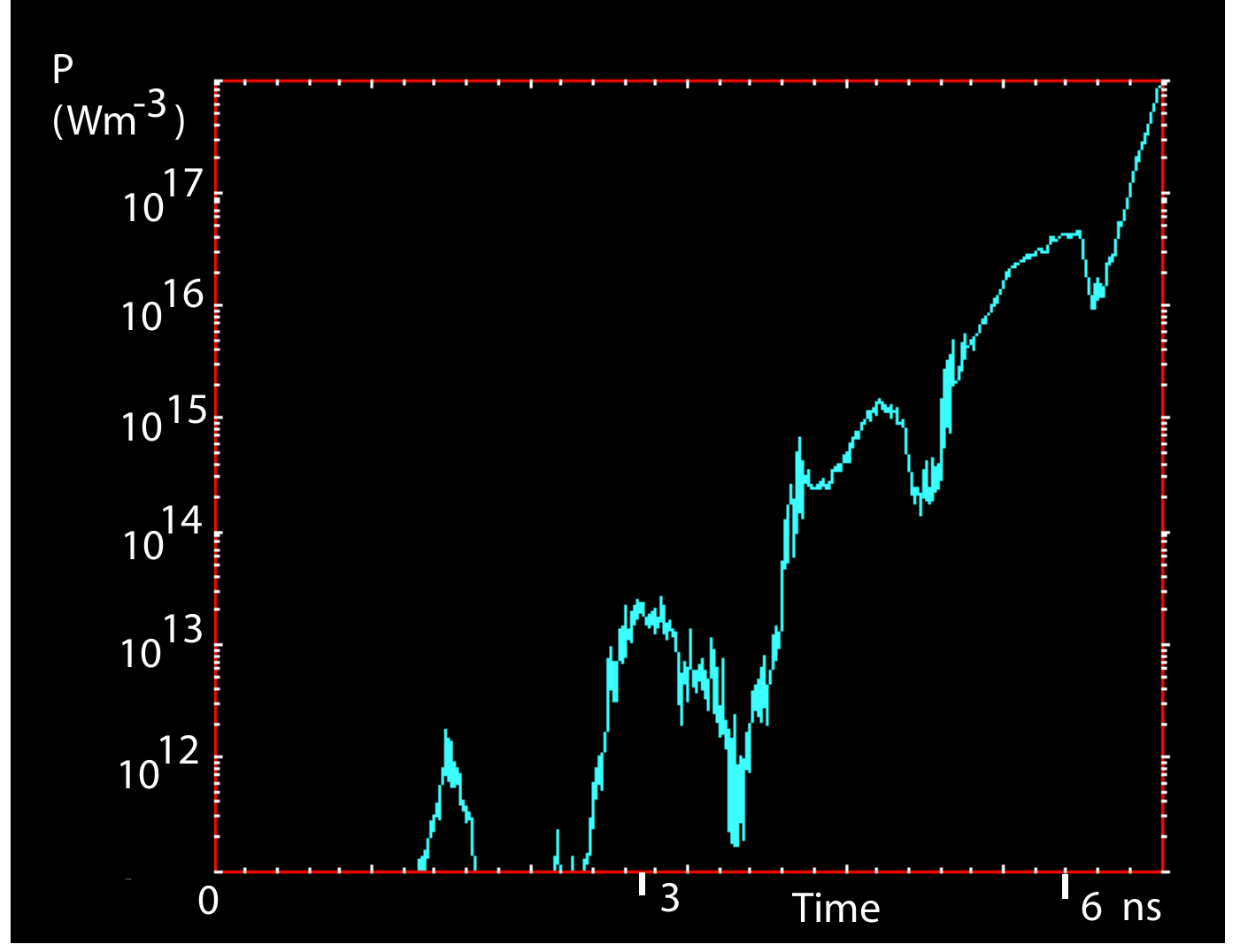}
\caption{The time evolution of optical radiation produced in the arc.}
\end{figure}

The primary source of heat into the surface is ion impact, since these ions would deposit all their energy of 10 to 100 eV/ion in copper in the top 1 nm.  The surface heating is modeled in OOPIC assuming all heating is due to incident electron and ions.  The surface heating is thus primarily a function of the plasma density and is proportional to it.  In Fig. 11 we show a typical plot of the rise of the surface heating (more precisely the heat flux for an arbitrary surface volume) with time, using a simple model involving heating only the top few atomic layers.

\begin{figure}  
\includegraphics[scale=0.38]{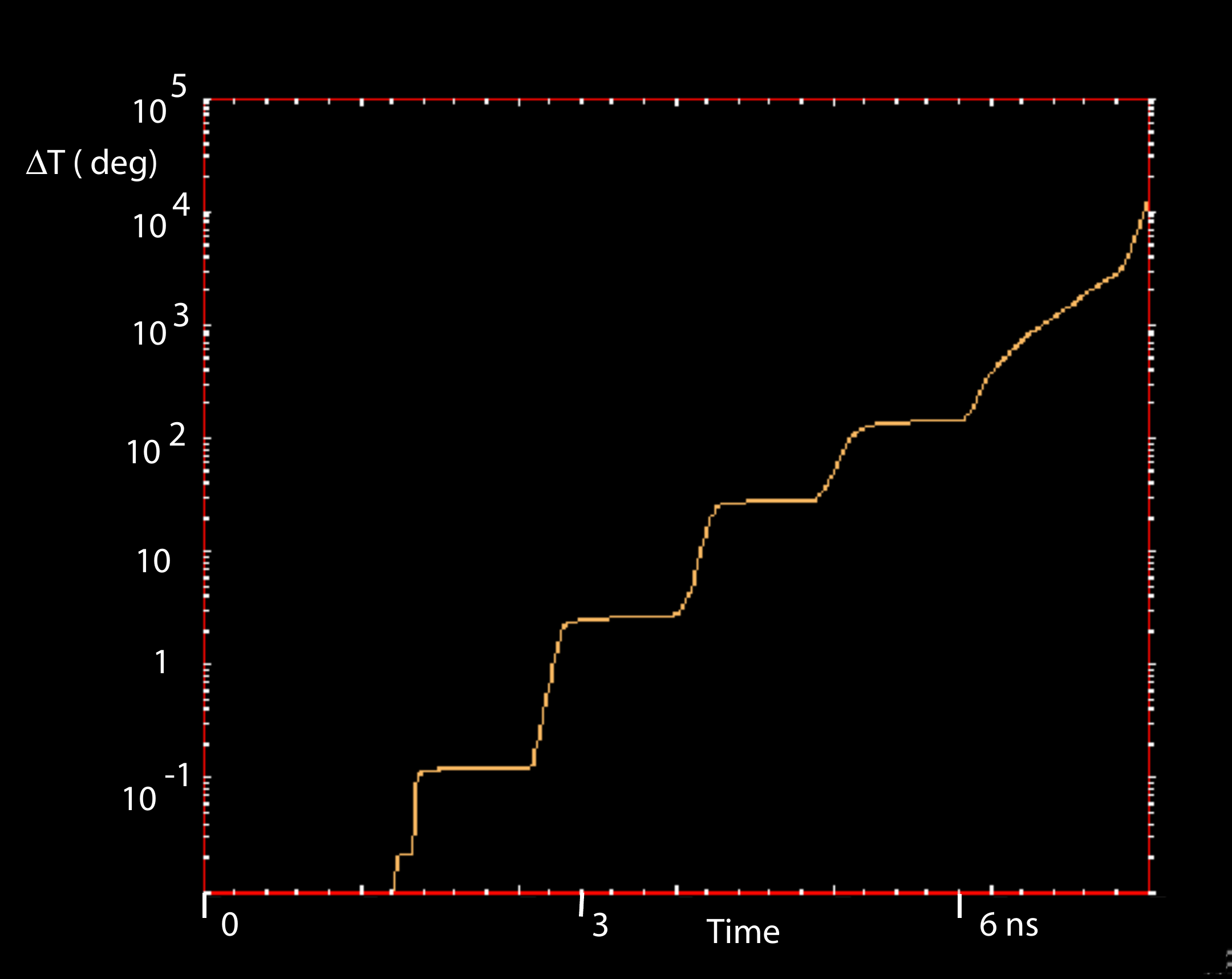}
\caption{Predictions from the plasma simulation of the surface temperature (or heat flux to the walls) of an element of the asperity, as the arc develops. }
\end{figure}

\subsection{Plasma pressure, local melting,  and particulates}

Simulations show that the arcs produced at these high gradients have a high plasma beta, $\beta_p$, defined for plasmas in thermal equilibrium as \[ \beta_p = nkT/(B^2/2\mu_0 ),\] where $n$, $k$, $T$, $B$, and $\mu_0$ are the plasma density, the Boltzmann constant, plasma temperature, magnetic fields on the order of 1 T, and permeability constant, respectively.  The plasmas are inhomogeneous, nonequilibrium, cold, weakly ionized, non-neutral, collisional, and inertially confined. When first formed, these plasmas have two weakly interacting electron populations. 

The acceleration of liquid droplets that splash into the walls of the structure is a common feature of cathodic arcs; see Fig. 22 of Ref. \cite{andersbook}.  Although these fairly large liquid drops could have some charge, they move so slowly (a few meters/second) that they would not see, or be affected by, the rapidly fluctuating rf fields and thus must obtain all their acceleration from the local plasma.  The generation of these particles is discussed at length by Anders \cite{andersbook}.

\subsection{Cavity shorting currents}

The major energy loss mechanism for the cavity should be the flux of plasma (field-emitted, ionization, and secondary emission) electrons that are accelerated to relativistic energies into the opposite wall by the cavity fields.  These electron currents should exceed a few amps, limited by space charge, and be accelerated to energies of a few MeV, depending on the geometry of the cavity, thus amounting to power levels of a few MW. 

Accelerator cavities operate at average gradients up to $\sim$100 MeV/m.  The cavity resonance then produces B fields on the order of $B=E/c\sim 0.3$ T, where $c$ is the speed of light.  Since the relativistic electrons are produced around $E_{max}$, the field is changing at a rate of $dB/dt = B\omega$, and we have shown in the previous section that the plasma may produce electrons for approximately half an rf cycle, roughly $\delta t = 2/\omega$.  As a numerical example, 4 MeV electrons have a rigidity of about 0.2 Tm and might be spread over a length, giving $dr \sim 1$ cm.

We have shown in Fig. 2c that, experimentally,  breakdown events in our pillbox cavity produce fluxes of about 4 MW of relativistic electrons to the opposite wall of  the cavity.  Modeling shows that these electrons are emitted from the plasma very close to the surface.  Since the potential difference between the walls of the cavity $E_{max}$ is about 4 MV, these relativistic electrons represent currents on the order of a few amps.
 
The thermal pulse $\Delta T = Q/Cm$, where Q is the cavity energy, $C$ the specific heat, and $m$ the mass of the wall being heated, should be easily detectable, either through the  low-energy bremsstrahlung that depends on the energy of the electrons \cite{emshower}
or later when heat has diffused into a larger volume of the wall.  

The deposited energy per unit volume may be lower at the surface and increase with depth as the electron gives energy to secondary particles.

During the arc, OOPIC Pro simulations have shown that electrons are accelerated from the arc over almost half the rf cycle because of the large gradient at the surface.  As the direction of the electric field changes, these electrons will be accelerated back toward the arc plasma, providing an additional heating into the arc region.

The special case when there is a strong magnetic field parallel to the electric field is  interesting.  Under these conditions the electrons and ions are ``pinned" to magnetic field lines, and both field-emitted electrons from normal asperities and the intense, relativistic beams from breakdown plasmas are forced to deposit all their energy in a smaller volume of material \cite{PR1,PR2,palmer}.    The normal field-emitted beams have been detected with Polaroid film and glass slides, and the much more intense beams from damage spots from breakdown events have been seen on the cavity walls and on the titanium vacuum windows, one of which burned through the metal to produce a vacuum leak.  These are described in Ref. \cite{PR1}.  Both the breakdown and field emission spots are circular; however, field emission beams are better measured because they can be predictably produced.

\section{non-Debye plasmas}

Classical plasmas are defined by three conditions: (1) the dimensions of the plasma, $L_p$, must be larger than the Debye screening length, $L_p > \lambda_D$; (2) the number of charges in a Debye sphere, $N_D$, must be much larger than 1; and (3) the plasma frequency, $\omega_p$, must be greater than the collision frequency,  $\omega_c$ or, $\omega_p > \omega_c$, so the plasma responds to external forces electromagnetically rather than collisionally \cite{chen}.   While these conditions are generally met in the early stages of the discharge, the plasma looks less and less classical as the arc evolves.  This evolution complicates modeling, because most codes assume that the conditions for a classical plasma (two body collisions, classical sheaths, etc.) are met, and thus the use of these codes becomes less appropriate as the densities increase.  

The rapid density increase shown in Fig. 6 is due to the rapid ionization of neutrals and occurs at low plasma temperatures.  The density and temperature ranges seen by the plasma are shown in Fig. 12, showing the Debye length of the plasma and the assumed surface field if the plasma potential remains at 75 V, as shown by the OOPIC results.  As the Debye length gets shorter, the number of particles in the Debye sphere becomes smaller, ultimately limiting the development of the sheath potential at the metallic boundary.  Moreover, as the Debye length decreases, the surface field will also increase as $\phi/\lambda_D$, ultimately reaching very high values, causing increased field emission, and perhaps ultimately reaching the space charge limit.  The relative density of neutrals is not known, so the collisionality of the plasma cannot be determined.

The OOPIC simulations show how the space charge limit appears in PIC codes.  Figure 9 plots $v_z$ vs $z$ for both trapped electrons and field emitted electrons.  At the space charge limit the local potential near the surface oscillates so that electrons are emitted in bursts, causing a reversal of the electric field near the surface.  In time, most of these electrons are pulled into the plasma; however, the field emission currents are highly modulated.  The time dependence of space charge limited emission is not generally considered.

\begin{figure}  
\includegraphics[scale=0.45]{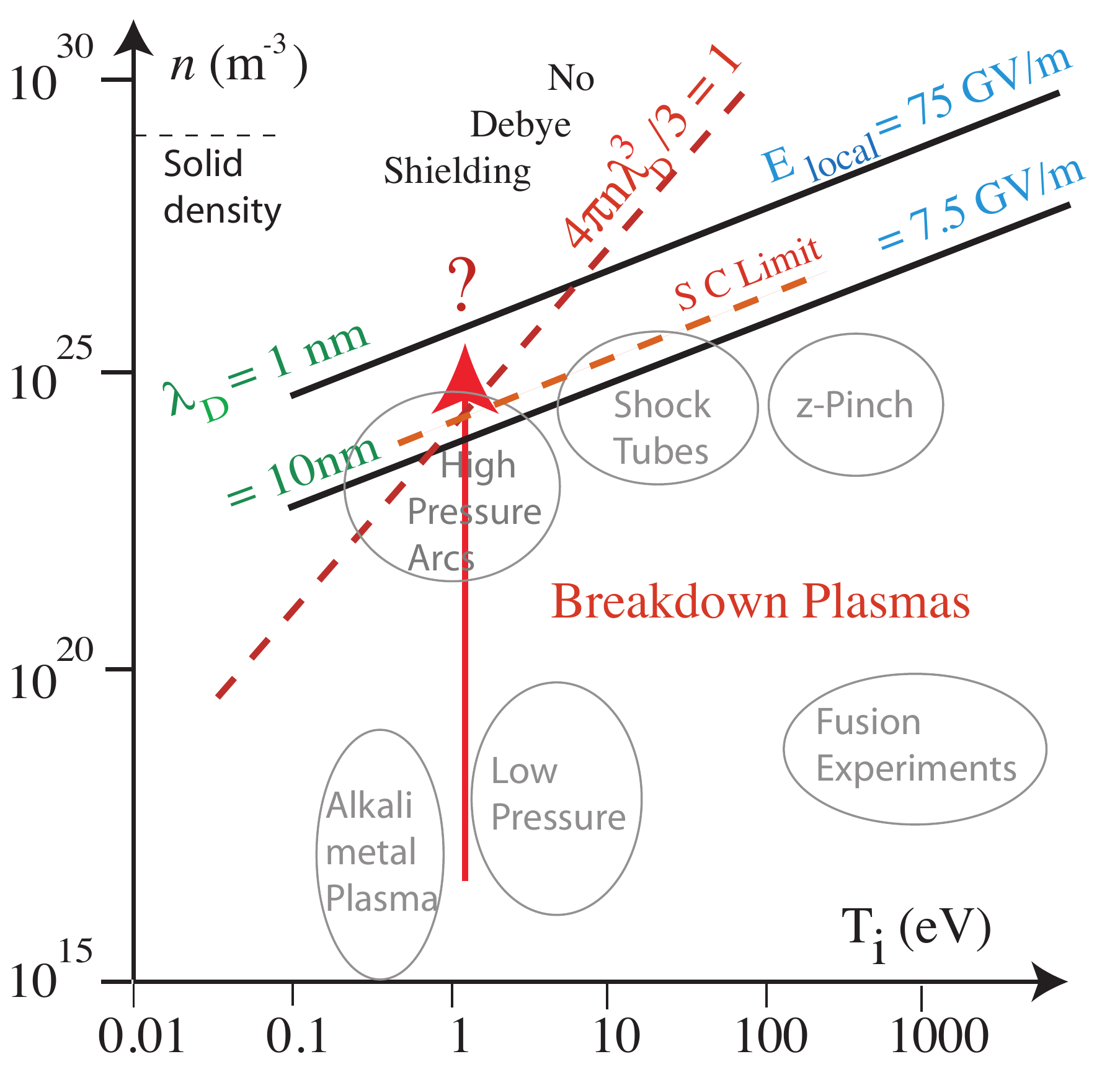}
\caption{The plasma parameters obtained from the OOPIC simulation plotted against the plasma density and ion temperature, showing the constraints of the plasma, the number of particles in the Debye sphere, $N_D$, the Debye length, $\lambda_D$, the approximate surface field assuming a plasma potential of 75 V, and the space charge limit for field emission.  The tic mark on the line shows the approximate surface field obtained from Tonks-Frenkel analysis.}
\end{figure}

\subsection*{Unipolar arcs}

While the properties of dense plasmas in contact with materials are not well understood theoretically, the properties of unipolar arcs may provide useful experimental information. Unipolar arcs were first described by Robson and Thonemann, but more recently in the context of laser ablation and tokamak limiters by Schwirzke .\cite{unipolar,Robson,schwirzke,unipolarW}.   A unipolar arc is a dense plasma in contact with a large metallic surface These arcs can be produced in a large number of ways.  A recent workshop at Argonne has summarized many aspects of the theory and applications of unipolar arcs.  We find that the best references on this mechanism are papers by Schwirzke et al. \cite{schwirzke,schwirzke1}.  A general picture of the unipolar arc mechanism is shown in Fig. 13. We consider the possibility that the unipolar arc could be inherently unstable. 

Unipolar arcs have been discussed at length in a variety of environments, and these arcs seem to be an important mechanism in determining the limits to accelerator gradients, tokamak stability, and efficiency. At present, however, little effort is devoted to understanding how these arcs function and what the defining parameters are.

Once arcs form, a number of mechanisms might cause them to terminate spontaneously and move to a new location. For example, (1) if the sheath disappears due to a non-Debye plasma, this might disrupt the plasma; (2) growing neutral density might produce excessive losses and cool the plasma; (3) excessive radiation loss might affect the arcs' behavior; (4) space charge cutoff might disrupt electron flow; or (5) surface melting might remove efficient emitters.  OOPIC simulations based on the mechanisms itemized in Fig. 5 have shown that as the ion density increases, the excess net charge density,  
\[ n_n = (n_i - n_e) \sim10^{20} /m^3,\] 
remains roughly constant as the arc develops.  A significant component of the net surface charge is due to the field-emitted currents, which are variable, even at comparatively low surface fields.  As the surface electric field under the plasma rises to values on the order of a few GV/m, it becomes possible for the whole surface area under the plasma to field emit. At a field of 5 GV/m with a current density of $10^6$ A/m$^2$, this current would amount to $6 \times 10^{24}$ electrons/m$^2$/s, capable of shorting out the plasma charge in 1 ns.  Since the field emitted currents go like $\sim E^{14}$, higher surface fields could cancel out the net plasma  charge in a much shorter time, essentially turning the arc off locally.  In practice, the plasma should be large enough that the arc could re-establish itself elsewhere.  This mechanism could explain the ``chicken-track" pattern of damage in some structures.  It is also consistent with the quantization of arcs described by Mesyats \cite{Mesiats}, although the mechanism would be different from the ohmic heating of liquid metal that results in ``ectons," which seem to define when the metallic jet that drives the plasma, in that model, burns itself out.

Once formed, these unipolar arcs can maintain themselves for long times, evidently deriving energy from their plasma environment or stored energy.  Kajita describes an analysis of the trail of arc pits left by these arcs after an arc was excited by laser ablation \cite{kajita}.

\begin{figure}  
\includegraphics[scale=0.3]{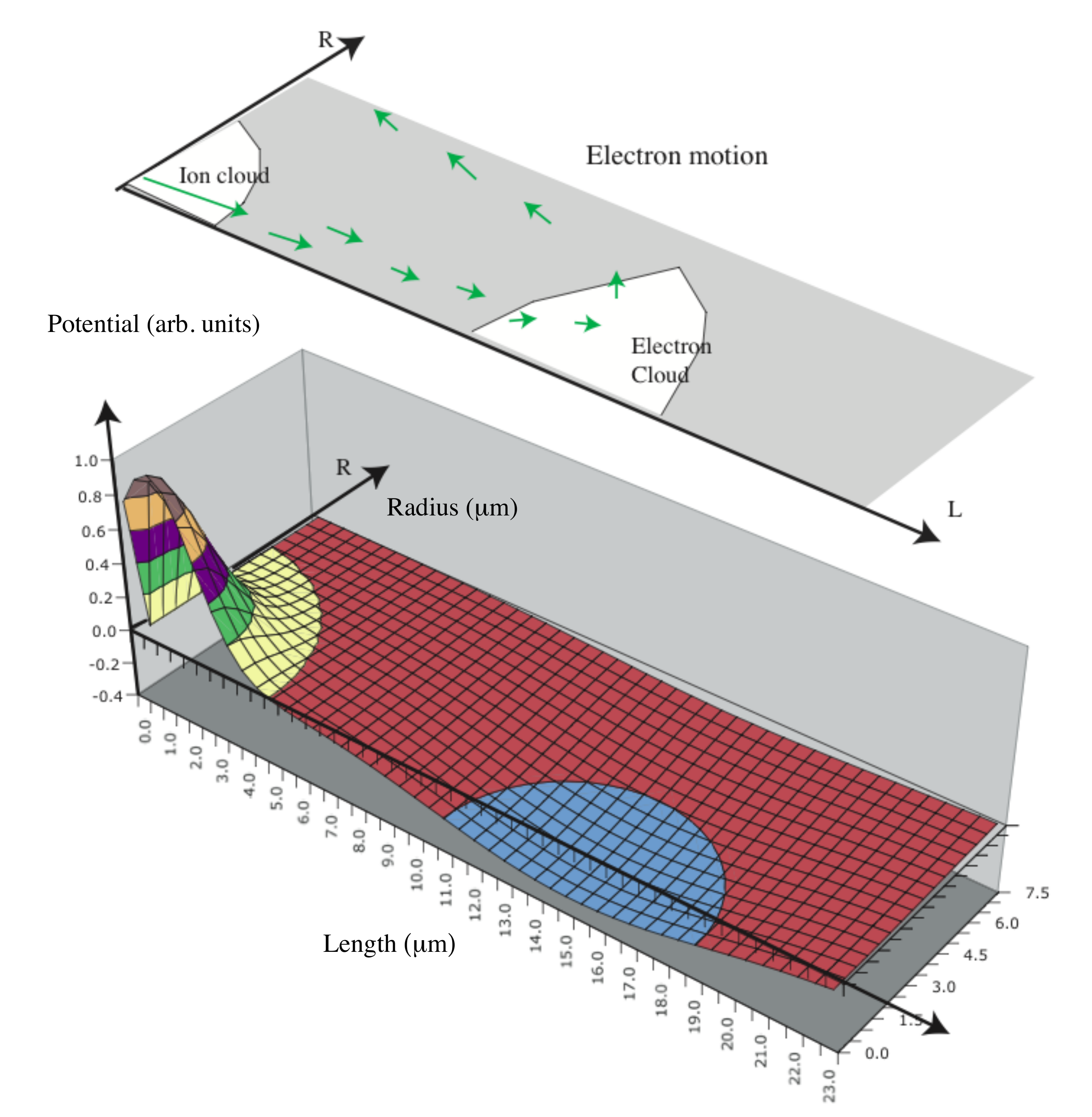}
\caption{General picture of the unipolar arc, with electron motion driven by the electric fields.  Electrons are pulled into the plasma by the surface fields; however, both the electrons and ions then diffuse away, their relative velocities resulting in a net positive charge at the center of the arc.  It seems consistent with the references to assume that the electron current driven by this potential could be a transient at any given location.}
\end{figure}

\section{Surface Damage}

A wide variety of surface damage due to arcs is seen in rf structures, and an even wider range of damage is seen in other arcing environments.  We think it is useful, however, to consider two extreme cases of arcs in rf systems: (1) killer arcs, which can short out the electric field of the structure and quickly remove the driving field from the arc, and (2) parasitic arcs, which occur in comparatively low electric field regions of the cavity or other structure and are able to operate for comparatively long times without interfering with the driving field.  The prototypes for killer arcs are arcs at the irises of linac rf structures that can completely remove all the energy in the structures in a few rf cycles, leaving small pits.  The prototypes for parasitic arcs are the arcs that leave extensive tracks on the walls of tokamaks, after evidently burning more or less continuously for timescales of $\mu$s to seconds.

In rf systems, surface damage is produced by two causes: (1) the electron beam that traverses the cavity and (2) the arc that develops close to the wall.  We assume that the damage produced by the electron beam is described by the electron kinetics and that the damage produced by the arc is described by a process essentially similar to the unipolar arc \cite{schwirzke}.  We believe that the damage caused by the arc is usually the dominant effect.

The  high local sheath potential and the distant sink for electrons leaving a small arc volume are the defining properties of the unipolar arc, studied extensively in the 1970s and 1980s in tokamak and laser plasmas.  Unipolar arcs are described by Schwirzke as a ubiquitous discharge that occurs between a plasma and an electrically conducting surface driven only by the sheath potential that exists between the plasma and the wall  \cite{schwirzke,Robson,unipolar, unipolarW}.  These arcs were studied not only for their basic interest as a unique phenomenon but also for their relevance to the production of metallic contaminants in  otherwise clean plasmas.  The unipolar arc mechanism efficiently converts plasma energy into surface damage, and this damage is primarily in the form of craters of various sizes.

\subsection{Self-sputtering}

While plasma initiation requires an initial source of atoms or ions above the surface to couple power from electric fields to the plasma energy, maintaining the plasma requires self-sputtering.  Kajita and others \cite{kajita,schwirzke,anders1} have shown that the arcs, once started, will maintain themselves for a significant time, although not always at the same location. While the duration may not be particularly important in rf arcs, this process is required for a complete picture of arcs in a variety of environments.

As shown in a number of papers \cite{insepovsput, anders2}, the development of a self-sustaining arc depends on the self-sputtering coefficient being significantly larger than 1.  We have explored the self-sputtering coefficient for copper ions as a function of surface temperature at or above the melting point and for high electric surface fields using molecular dynamics calculations showing that both high surface temperatures and high surface electric fields can independently increase the self-sputtering rate to greater than 10 either near the melting point of the material or at fields at or above 3 GV/m; see Figs. 11 and 12 of Ref. \cite{insepovsput}.  The dynamics of sputtering from liquid metals or sputtering from surfaces at high electrical gradients has not been extensively explored, so modeling these effects seems to be the most useful way to evaluate the effects.  Melting decreases the binding energy of atoms at the surface, and high electric field gradients induce surface charge that should pull atoms from the surface.  It seems reasonable to expect that the combination of both high fields and high temperatures would further increase the self-sputtering.  High self-sputtering rates produce fluxes of neutrals into the plasma that must either increase the plasma density or produce increased fluxes of ions that impact the surface, both processes ultimately increasing the plasma density and thus the surface field.  We show the dependence of self-sputtering on surface temperature and field in Figs. 14 and 15, respectively.

We assume that the primary damage mechanism is erosion caused by self-sputtering, with an erosion rate, in meters/second, that is effectively determined by the thermal current times the sputtering yield, 
\[rate \sim n_I \ v_{t}\    Y(\phi, \lambda_D, T) V_{Cu},\]
where $n_I$ is the density of plasma ions; $v_{t}$ is their thermal velocity $Y(\phi ; \lambda_d, T)$ is their self-sputtering yield for a given plasma potential, Debye length, and surface temperature, respectively; and $V_{Cu}$ is the volume of a copper atom.  The mean of the magnitude of the thermal velocity is expressed, in one dimension, as
\[v_z = \sqrt{2kT_i/\pi M} \]
and depends on the ion temperature $T_i$ and the ion mass $M$.  The intensity of these currents is described in Ref. \cite{andersbook}.

\begin{figure}  
\includegraphics[scale=0.5]{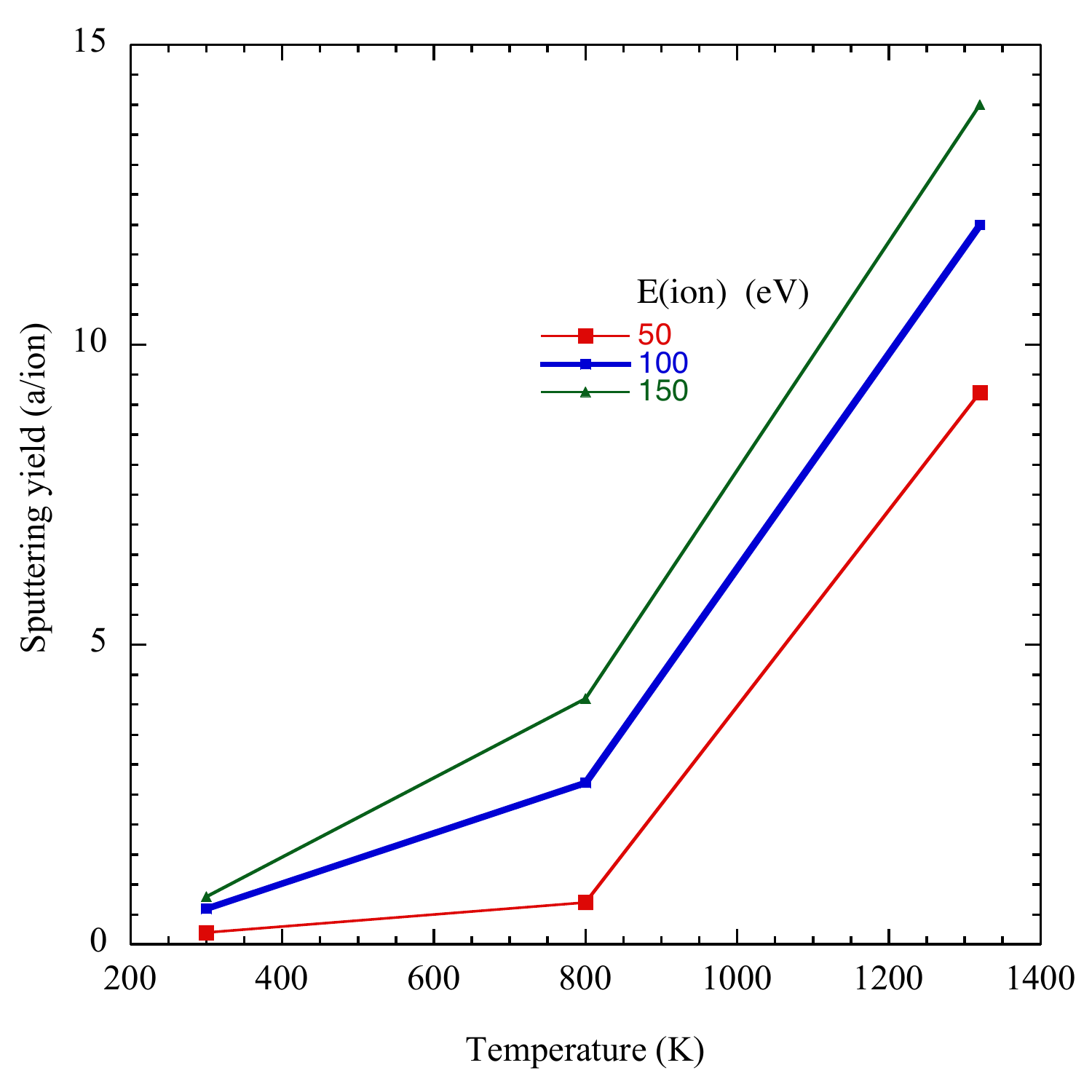}
\caption{Calculated sputtering yield as a function of T, showing the large increase at the melting point \cite{insepovsput}.}
\end{figure}

\begin{figure}  
\includegraphics[scale=0.5]{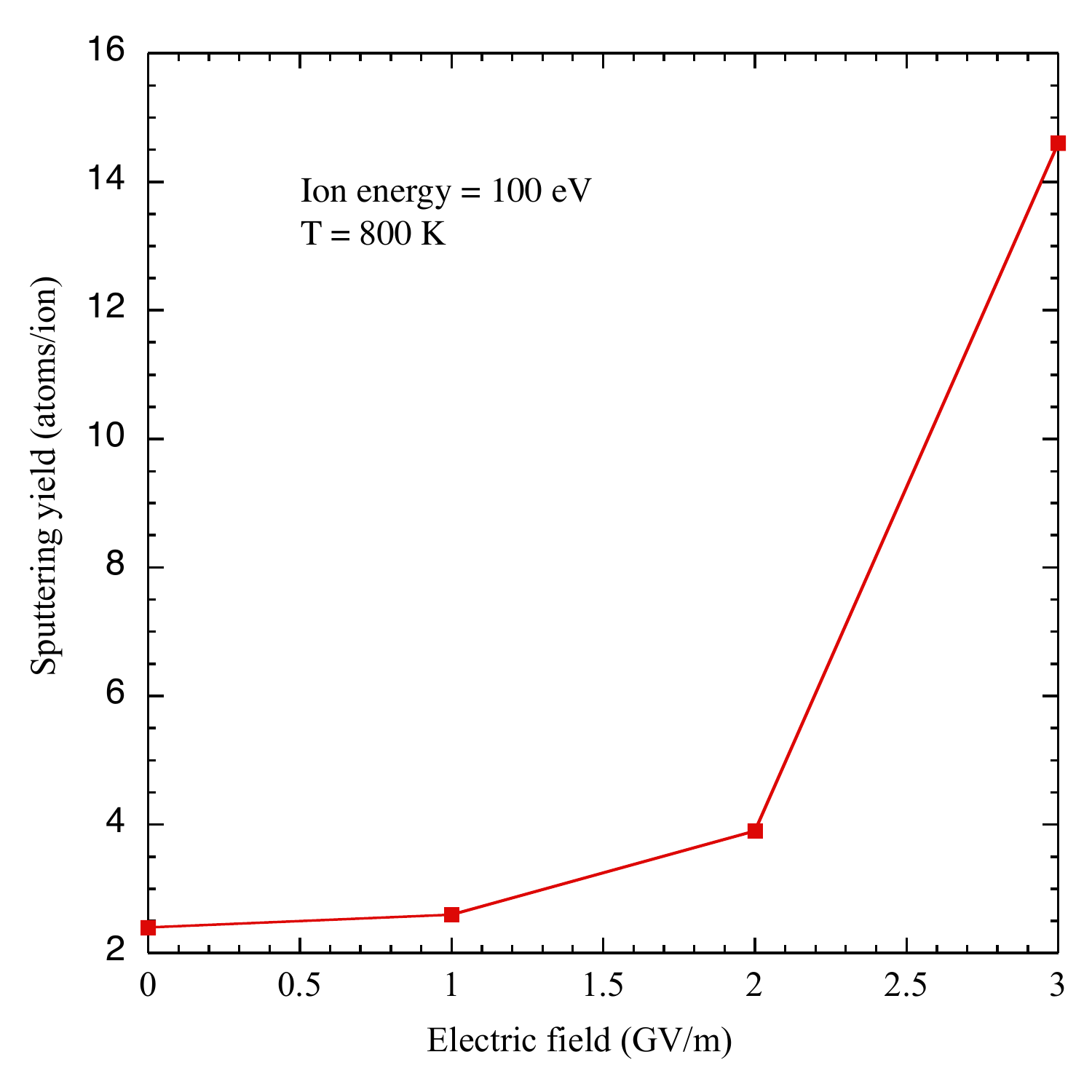}
\caption{Calculated sputtering yield as a function of E from molecular dynamics showing a large increase as the electric field increases \cite{insepovsput}.}
\end{figure}

\subsection{Ripples}

One of the fundamental problems in the study of arcs is precise measurements of the basic plasma parameters.  In the model described here, the surface field is perhaps the most important parameter because it drives the field emission currents and the ion currents.  Because these arcs occur unpredictably over large areas, last only a few nanoseconds, and have sizes of a few micrometers and because the surface fields are effectively screened a few nanoseconds into the plasma, direct measurements of the surface field are difficult.  Here we describe an indirect method of measuring the surface electric field by looking at the surface damage left by the arc, as shown in Fig. 16.

Formation of surface periodic and irregular structures such as ripples, cones, bubbles, or a combination of them has been observed on rf cavity surfaces that have encountered vacuum breakdown on the first walls on the tokamak chamber and on the surfaces irradiated by laser beams during hypervelocity impact welding.
Thus, this formation has been the  subject of detailed studies of experiments and computer simulations.  Indeed, understanding the ripple and microstructure formation is important for all these research fields and can be used, we believe, to estimate the range of surface fields at the center of an active arc.

At least two types of surface ripples are known.  The first type is formed on an ion-bombarded and heavily eroded surface, either conducting or insulating.  The crests of such type of ripples are oriented perpendicular to the direction of incoming ion beam, so that the wave vectors of the 2D surface wave structure is oriented parallel to the ion beam.  The direction of the ion beam should be inclined to the surface, to be able to create ripples (since normally oriented beams do not create ripples).  These effects are described by the Bradley-Harper model \cite{BradleyHarper,Chason,carlos}. 

Impact of clusters has also been seen to produce surface ripples in gas cluster ion beams \cite{insepovyamada}. 

The second type of ripples is similar to capillary waves on a liquid surface and can be created on a conductive surface in a strong surface electric field.  The appearance and dynamics of capillary waves under a wide range of conditions are well understood \cite{he}.  These structures can be driven by a number of perturbations, and there is an extensive literature on the subject.  This field is based primarily on work done by Tonks \cite{tonks} and Frenkel \cite{frenkel} in the 1930s, where the basic dynamics and stability were worked out.
Our interest in this phenomenon is related to the possibility that these structures can help describe the environment in the interior of dense plasma arcs.  

\begin{figure}  
\includegraphics[scale=0.7]{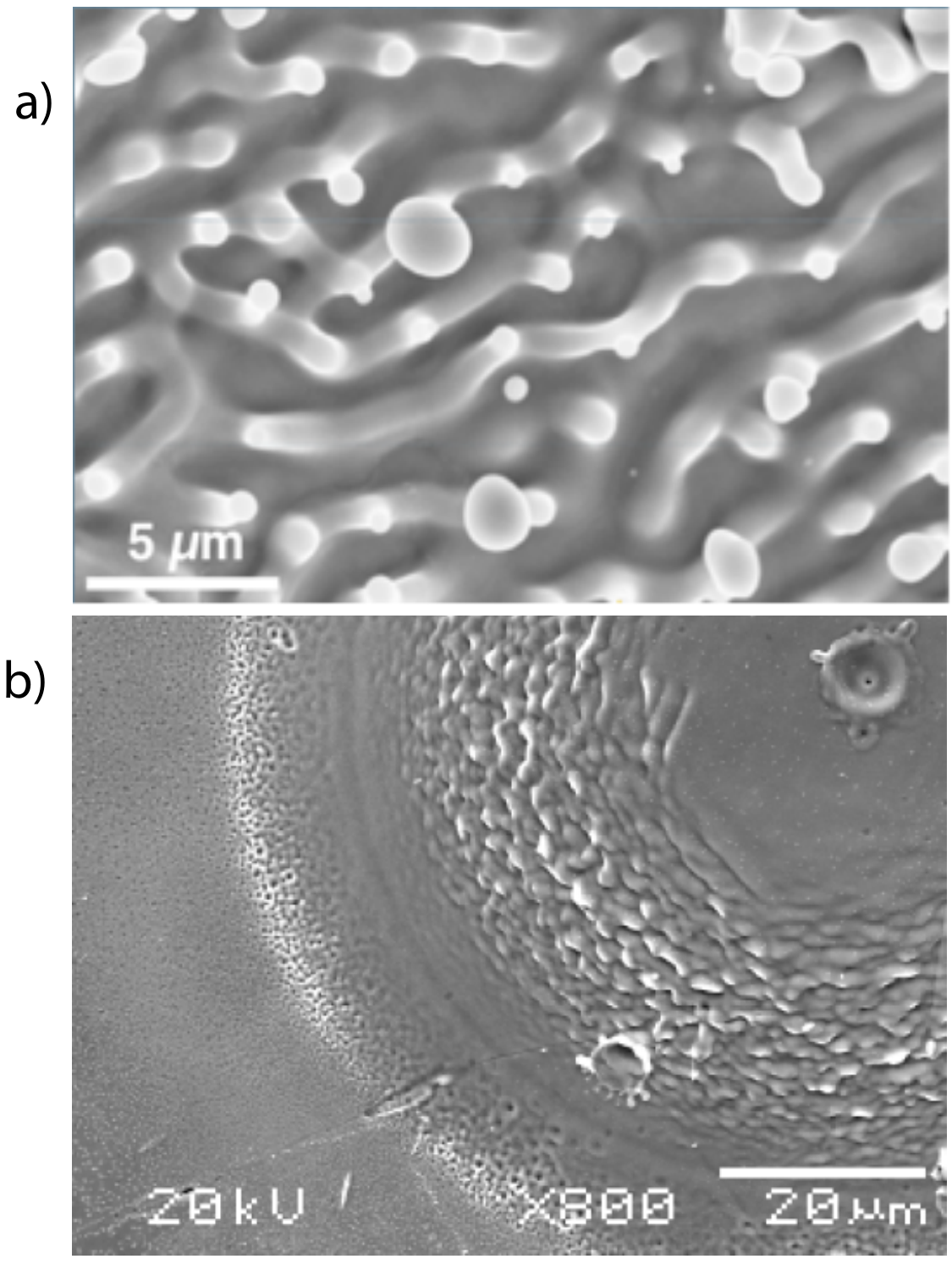}
\caption{SEM images showing ripples produced in arcing: (a) CLIC structure  \cite{Gonzalo}, and (b) high-power arc damage \cite{carlos}.  The upper image shows the sort of structure expected from capillary waves or spinodal decomposition; the lower picture shows structure consistent with low-angle ion bombardment.  }
\end{figure}

We look at capillary waves and microstructure present in the arc pits left after the arc is gone.  The properties of electrostatically driven capillary waves appropriate to our example have been described by He et al. \cite{he} and Prergenzer \cite{pregenzier} for different metals, by means of dispersion relations giving the wavelengths and growth times of the waves structures produced.   

\subsection{Experimental measurements of liquid metal surfaces}

Ripples and structures in liquid metal surfaces are an example of electro-hydrodynamic spinodal decomposition of a flat liquid thin films into structures and cusps (spinods) that can be either two dimensional or three dimensional \cite{he}.  The well-known phenomenon of Taylor cones is one example of these structures.  

The simplest method of understanding structures on the surface of liquid metals is to consider only the electrostatic tensile stress 
\[\sigma = \epsilon_0 E^2/2,\] 
where $\sigma, \epsilon_0$, and $E$ are the surface stress in N/m$^2$, the permittivity of free space, and the electric field, respectively; the surface tension, $\gamma$, for molten copper is approximately 1.3 N/m.  Equating the pressure due to surface tension and electrostatic forces, as is done to determine the dimensions of bubbles, we find that the equilibrium radius for spherical surfaces is  
\[r = 4 \gamma/(\epsilon_0 E^2)   \]
or
\[E=\sqrt{4 \gamma/\epsilon_0 r}.\]
This analysis neglects the pressure exerted by plasma ions, so the electric field determined in this way would be a lower limit.

A paper by Tonks addresses the problem of electrostatically driven effects on liquid metals \cite{tonks}.  In the examples we consider here, the electrostatic forces are orders of magnitude larger than the forces due to gravitational acceleration, so this term is always neglected.  Figure 17 shows simulations of the ripples that can be produced.  In the case of plasma arcs, the problem is complicated by the contribution of the plasma pressure due to ion fluxes that push on the surface, opposite to the electric tensile stress. Thus, we expect this analysis would give a lower bound to the expected electric field since it is difficult to evaluate the plasma pressure, which may vary widely from example to example.  Both the electrostatic and plasma pressure would be expected to be a function of the surface geometry to some extent, further complicating precise solutions.

Pictures of the surfaces of copper structures that have experienced significant arcing show a variety of morphologies in arc pits and larger areas;  however, many surfaces show considerable structure with characteristic dimensions around one micron,  as shown in Fig. 16.  These figures are examples of the variety of structure seen. 

The fields derived from the OOPIC analysis and confirmed experimentally by this method, roughly 1 GV/m or greater, are consistent with huge currents being produced locally by field emission.  We note that megagauss magnetic fields have been produced in laser-produced plasmas, where no net current is introduced, and the high local electric fields we derive are a simple explanation of this phenomenon \cite{stamper}.

\begin{figure}  
\includegraphics[scale=0.8]{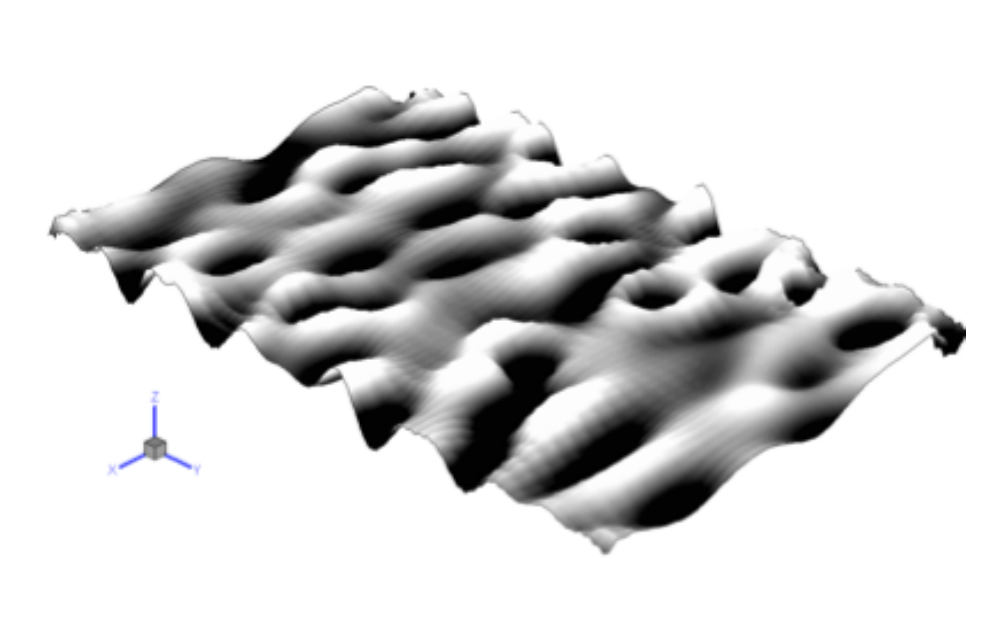}
\caption{Simulation of the Tonks-Frenkel instability based on a numerical solution of the surface dynamics Kuramoto-Sivashinski equation for a copper surface where the surface tension was modified according to the Tonks model by adding the negative electric field pressure $-\epsilon_0 E^2/2$ to the effective surface tension.}
\end{figure}

\subsection{Field enhancements}

A number of papers measuring the properties of the pre-breakdown asperities have implied that breakdown is triggered by high surface fields on asperities, but the shape of these asperities has been widely debated.  Defining a field enhancement, $\beta$ = (local field/average surface field), considerable data shows field enhancements in the range of 100--1,000 \cite{Latham,padamsee}.  The areas of the pre-breakdown asperities have also been measured, though with less precision, giving dimensions over a wide range, down to a few nm$^2$. 

In the early 1950s Dyke et al. carefully studied field emission and breakdown with tungsten needles and argued that surface failure was caused by ohmic heating of the surface due to Fowler-Nordheim field emission. However, breakdown occurred at the same values of the local surface field, $7$ to $10 \times 10^9$ V/m \cite{dyke}.  The problem of how local surface fields could be so much larger than the average surface fields on electrode surfaces has not been persuasively explained.  Calculations by Rohrbach showed how simple cylindrical geometries could produce large field enhancements \cite{rohrbach}.  This paper describes the field enhancements produced by a variety of simple geometrical shapes, finding that the fields at the tips of cylinders (whiskers) oriented perpendicular to the surface are enhanced by a factor roughly given by the length of the cylinder divided by its diameter .  Although cylindrical structures (whiskers, telephone poles, etc.) seem to be the most common explanation for field enhancements, structures fitting this description, with aspect ratios as large as 1000, are not seen, causing some confusion.

In the superconducting rf community, where field emission is a critical failure mechanism, it has been determined that particulates (ideally with sharp corners) were the primary source of the large enhancement factors seen in field emission, and examples of the Òtip-on-tipÓ model seem to explain both what is experimentally seen as the geometry of the emitters \cite{padamsee,safa}.

In addition to the micron-scale structure described above, we, and others, experimentally see a variety of sharp edges, corners, and cracks in the surface. This paper considers the field enhancements that would be present in these edges, corners, and cracks and how they might function as field emitters.  We describe numerical calculations of realistic geometries that produce high enhancement factors and are consistent with experimental data. We find that a variety of structures can produce the expected values of enhancement factors, although with small surface areas, so that a number of sources must contribute.  The electrostatic Laplace equations for these surface structures were numerically solved by using the finite-element multiphysics simulation package COMSOL \cite{comsol}.

\subsection{Evaluating field enhancements using COMSOL}

We show in Fig. 18a a SEM photograph at high magnification of the bottom of an arc pit (about 200 microns in diameter) from an rf cavity breakdown event showing cracks running in all directions. Because the crack junctions are numerous and topologically similar to the conical example discussed above, we have modeled the electrostatic fields assuming the tips of these junctions could be sharp.  

COMSOL is convenient for electromagnetic simulations in two and three dimensions [1].  It solves the Laplace or Poisson equations for arbitrary geometry by adaptive finite-element constructions where the smallest element size can be chosen to be at an atomic scale yet the system can be solved within a short computational time.  
We used an adaptive mesh structure;
a mesh is a partition of the geometric volume into small units of simple shape in the simulated system.  The Maxwell equations do not take into account the atomistic nature of the charges on the metal surfaces. 
Therefore, we set the minimum mesh size to 1 angstrom at the sharpest geometric locations, but this condition is not critical. The calculations were performed at different element sizes, and the results were obtained at the level where the dependence of the results on the cell size was negligible. The boundary conditions for the solution of the electrostatic problem were chosen as shown in Fig. 18. 

\begin{figure}  
\includegraphics[scale=0.8]{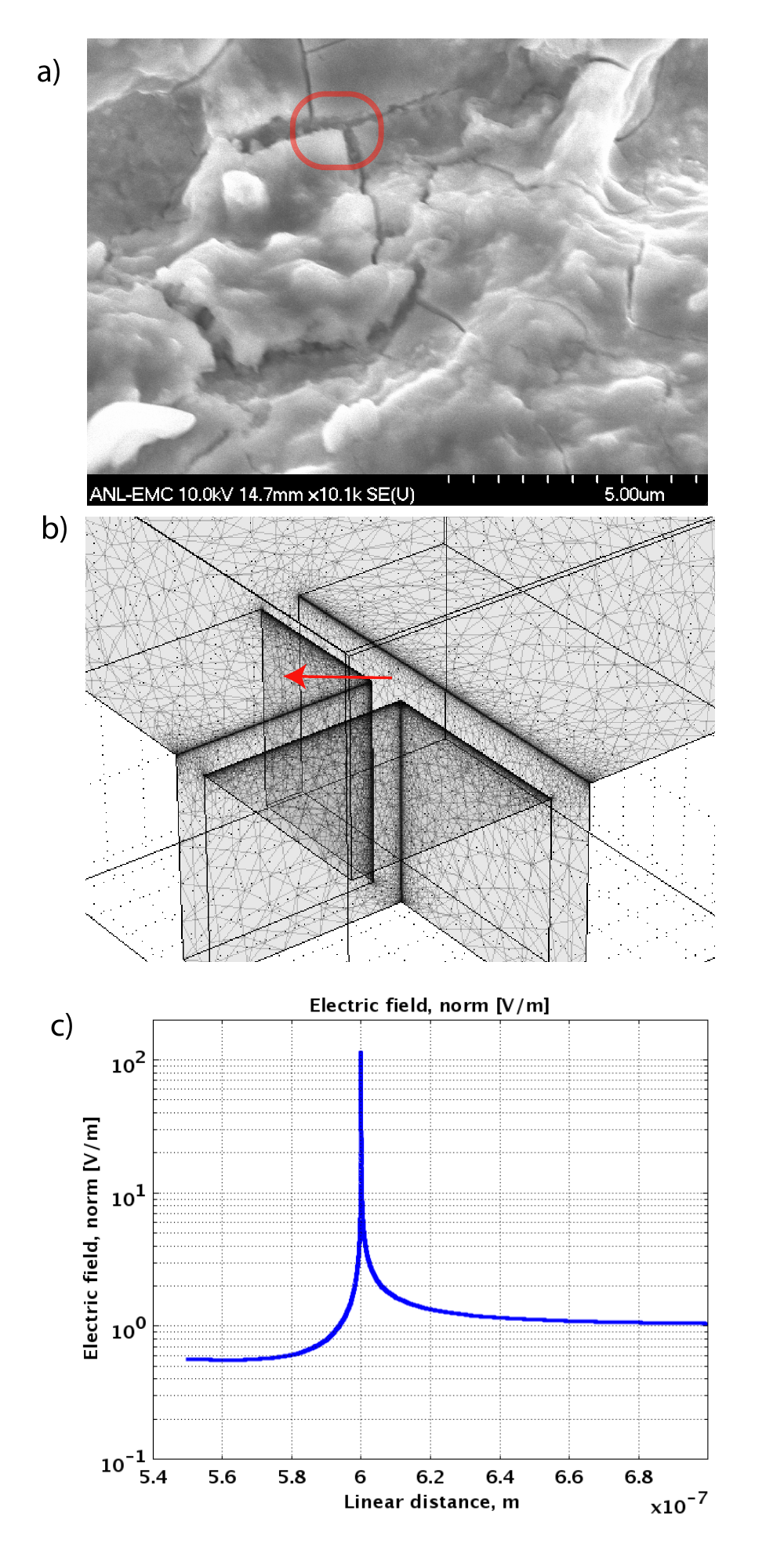}
\caption{ a) SEM image of arc pit cracks on the breakdown surface from an rf cavity after a breakdown event showing considerable microstructure.  The magnification on plot (b) shows our COMSOL simulation setting for a triple crack junction and (c) shows the calculated field enhancement at the triple crack junction, with $\beta$ = 140, which is close to experimental values in the range of $\sim$ 180--200 obtained in cavity measurements.}
\end{figure}

A two-electrode system was created, with the bottom electrode shown in Fig. 18b.  We have inserted sharp-edged cracks on the order of 300 nm wide in this electrode.  The top (flat) electrode was 2 microns from the bottom electrode.  The electric potential of the top electrode and the total distance between the electrodes, H, were chosen to generate the electric field 1 V/m everywhere in the system except near the tip's sharp spherical end.  Insulating boundary conditions for the side surfaces were applied in two- and three-dimensional symmetries. The voltage at the  top electrode is  2 $\mu$V, at a distance of  2 $\mu$, producing field enhancement along the red arrow in Fig. 18b, directly from the plot shown in Fig. 18c. The maximum number on the plot corresponds to the maximum field enhancement. 

Experimentally, high-magnification SEM pictures of arc pits show cracks and pits with a variety of sharp edges at the limit of the resolution of the microscope.  Mueller et al. have also shown SEM pictures of various physical shapes that produce high enhancement factors \cite{mueller,mueller2}.  We assume that these sharp edges are the source of the experimentally seen field enhancements and thus the field emission, and we calculated their properties.

We also calculated enhancements from conical structures (not shown).  The enhancements calculated for conical and sharp-edged geometries depend on how the singularity at the tip is handled.  We have compared sharp and rounded geometries and found that the rounded geometry essentially truncates the enhancement at a value comparable to where the cone radius is equal to the radius assumed for the tip.  We are interested primarily in the tips produced at triple crack junctions, however, since these seem to be numerous in SEM images.  While the emission from one of these tips would have insufficient area to produce the observed field emission currents, adding the contributions of many of these tips could produce the few nm$^2$ produced by fits to the field emission currents.  As seen in the SEM photo, the surface in which these crack junctions is imbedded is not flat and would be subject to a tip-on-tip enhancement, which could be significantly larger than the simple factor calculated for a flat surface.

The numerous cracks in this example can be explained by the rapid thermal cooling of the surface after an arc.  The melting point of copper is 1357 K; thus, molten metal in arc pits must cool about $1000\,^{\circ}\mathrm{C}$C before it reaches room temperature.  If we assume that arcs heat only the top micron of the material, the copper should contract by an amount 
\[ dL/L = \alpha \Delta T =  (17 \times 10^{-6} \times 1000) = 0.017.\]
Thus, roughly 2\% in any linear direction in the hot center of an arc pit should be a crack.  The appearance of cracking is a function of rapid heating and cooling of the surface layer.  In our samples we see cracks of width about 0.2 microns and about 10 microns apart.  We expect these crack junctions would produce large numbers of more or less identical, atomically sharp emitters.  We assume that these emitters, combined with additional enhancements from the local surface structure, would explain the spectrum of field enhancements seen in Fig. 3.  We have found that altering the angle at  which the cracks meet slightly changes the field enhancement; but the constraint that the surface must be flat, to first order, determines  that the emitters are similar.  We assume that all surface structures with high enhancement factors can be field emitters or breakdown triggers and are relevant to this paper. 

The heating produced by the arc may be able to locally melt (and round-off corners) in the region surrounding the arc center, which may avoid having larger areas being covered with emitters and the emitter density diverge.

These cracks and crack junctions are probably not the only defects with high field enhancement capable of producing breakdown.  Decades of optimization of the surfaces of superconducting rf structures has shown that it is extremely difficult to eliminate field emitters from these structures to permit them to operate at surface fields above 70 MV/m.  The implication is that a significant number of sites must have field enhancements on the order of 50 to 100, even in the cleanest systems.  When strict rules for particle contamination are loosened, one would expect significantly more contamination, field emission, and breakdown sites.  It is these sites that we expect would be removed with the initial conditioning.  Nevertheless, in a fully conditioned cavity, we expect these cracks to be the dominant cause of breakdown events.

\subsection{Ohmic heating by field emission currents}  

Following Grudiev et al. \cite{dyke} and Dyke et al. \cite{CERN4} scientists have assumed that vacuum breakdown was triggered by ohmic currents caused by field emission essentially exploding an asperity with the geometry of a whisker.  Cracks and corners with geometries similar to those shown in Fig. 18 are much more difficult to heat for two reasons: (1) the heating, which goes like the current density squared, is confined to a much smaller volume near the point of the corner and drops off very quickly from the point, and (2) the thermal mass is determined by the dimensions, $R_{th}$, of the volume accessible by thermal diffusion in one rf pulse.  This thermal mass is determined by the thermal diffusivity of copper, $D_{th}$,  $R_{th} \sim \sqrt{D_{th}t} \sim 0.3\  \mu$m, with $D_{th} \sim 1.1\times 10^{-4}$ m$^2$/s, and $t \sim$ 1 ns \cite{PR1}.  Since the volume of the thermal reservoir in equilibrium with the ohmically heated volume increases faster than the applied heat, $V_{th} \sim R_{th}^3 \sim t^{3/2}$, we show that the heating from a train of rf pulses would quickly saturate.   Combined with the large difference in resistivity between the tungsten needles used by Dyke et al. and the copper structures used here, heating in the corners of cracks should be much smaller that in tungsten needles.  Essentially, the resistively heated volume is measured in nm$^3$, and the heat sink (on the timescale of ns) is on the order of 0.1 $\mu$m$^3$, with a cooling time of a few femtoseconds.  

We have used COMSOL to evaluate the heating, assuming the apex of the field emitter is 9--10 GV/m and the field emission goes like $E^{14}$, giving a duty cycle of about 0.1. With the geometry in Fig. 19, the results show that the deposited heat is confined to a very small region near the apex of the corner and that the maximum thermal excursion is about 10 degrees per rf cycle with currents of $\sim 10^{12}$ A/m$^2$, for rf frequencies on the order of 1 GHz (depending on the exact geometry of the heated volume), which is not sufficient to produce significant heating.  Precise numerical calculation is difficult because of the range of dimensions, which ideally must include corners with dimensions of nanometers as part of a geometry with micrometer-sized structures, times from a few picosecond heat pulses applied over many microseconds, and many orders of magnitude in temperature, with large temperature excursions on a very slow overall heating profile. The rate of heating of the apex over long pulses would be affected by the volume of the heat sink, which we have estimated as (300 nm)$^3$, so the temperature rise would be on the order of 10$^O$(1/300)$^3$ or $\sim 10^{-6}$ degrees/ns.  

\begin{figure}  
\includegraphics[scale=0.3]{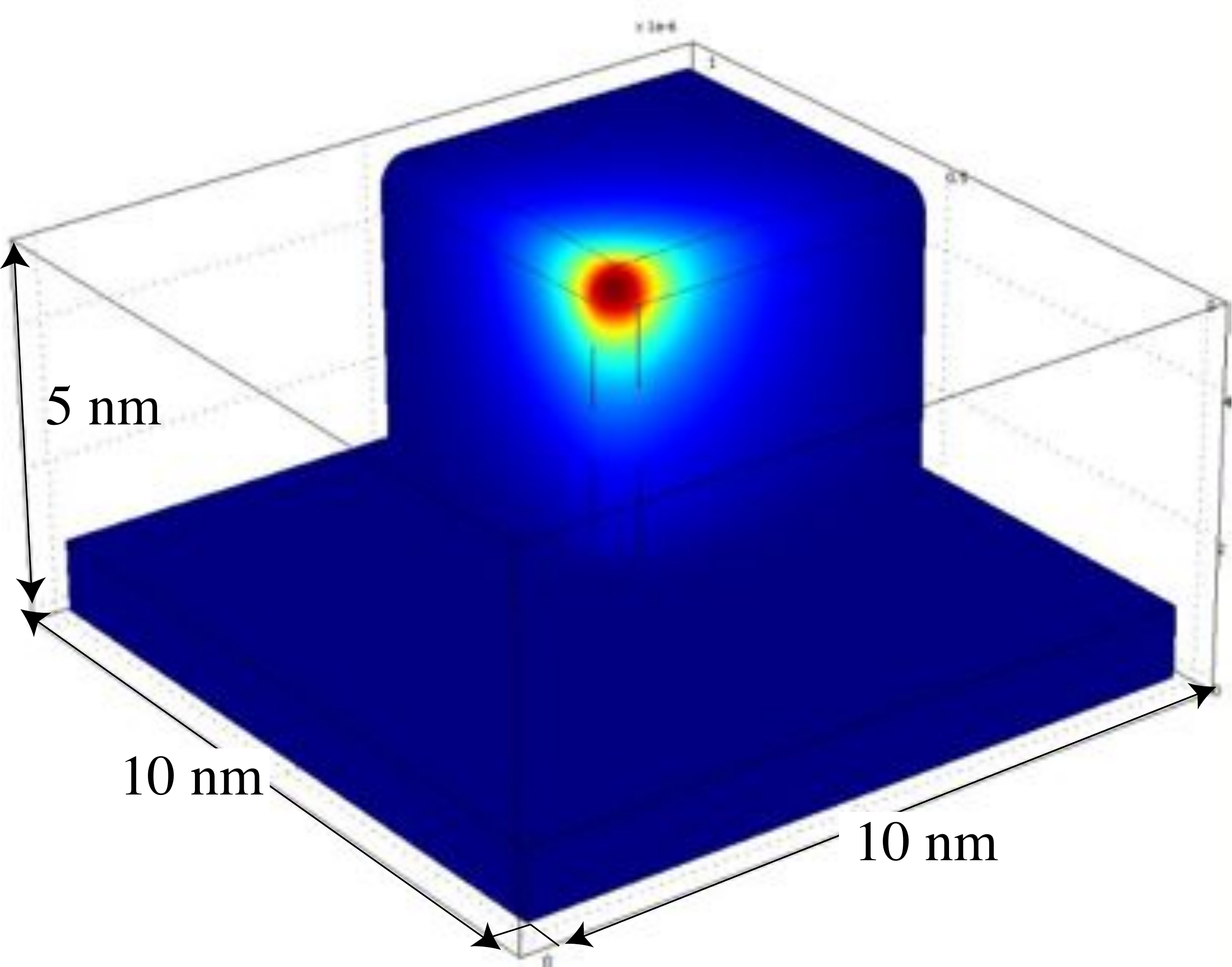}
\caption{Simulation of ohmic heating at the high field point of a 90-degree crack junction, (see Fig. 18) using a simplified geometry. A current density of $10^{12}$ A/m$^2$ is applied to the corner for 100 ps, giving a maximum thermal excursion of a few degrees.  After the current is removed, this heat rapidly diffuses and is undetectable 1100 ps later when the next current pulse starts.  }
\end{figure}

\subsection{Fatigue, creep, field evaporation, and electromigration}  

A number of physical mechanisms affect the surface of highly stressed structures, triggering a failure of the surface. As mentioned in Section I, the environment of the field emitter/breakdown site is  close to its ultimate electric field limits because of field evaporation, which tends to dull or smooth asperities, and a number of other effects, including fatigue, creep, electromigration, and ohmic heating---all of which can make asperties sharper.  Since these effects all depend strongly on the local electric field, which is dependent on a local geometry that is also not well known, it is difficult to evaluate the relative strength of these effects over the appropriate range of surface fields.  It becomes useful, however, to look at the dependence of individual mechanisms on the electric field, since the rates for these processes decrease rapidly as the field is reduced from its maximum value.

Field evaporation has been discussed at length in many references (see, e.g., Chapter 2 of \cite{milleretal}).  This effect is due to the electric field pulling individual atoms out of the potential well of the material.  The flux produced depends strongly on the electric field, with data showing field dependencies from $E^{50}$ to $E^{150}$, the high exponent leading to preferential erosion of more exposed atoms, producing a general smoothing effect.  For most smooth surfaces the surface fields at which this process occurs are a few tens of gigavolts/meter; however, these surfaces are unlikely to occur naturally in real structures, so the effect would likely be significant at lower fields for surfaces with some micro-roughness.

Fatigue and creep are active at the atomic level,, allowing strain in solid materials to respond to cyclic stress \cite{fatigue}.  Since stress is proportional to $E^2$, we show that this mechanism might be stronger farther from the ultimate materials limit than effects with higher exponents.  For creep, it is customary to consider the ``creep exponent" $n_C$, where
\[d\epsilon/dt = C \sigma^{n_C}e^{-E_c/kT},\]
where $\epsilon$ and $\sigma$ are the strain and stress, respectively;  $E_c$ is the activation energy for the creep mechanism; and $e^{-E_c/kT}$ is the thermal dependence of the process.  For very low failure rates we expect high values of $n$ \cite{fatigue}.  Any measurements of the creep exponent must be done over a fairly narrow range of failure rates. High rates would imply low exponents; low failure rates would imply operation near the failure thresholds and high values of $n$.

Electromigration describes how high current density electron flows and high temperatures can cause surface ions and atoms to move in the direction of the electron current \cite{electromigration}.  This phenomenon is common in integrated circuits where small gaps, high fields, and  high currents are present.  Antoine has pointed out to us that the environment of a breakdown site may be susceptible to this mechanism because of the high current densities moving near a field emitter and the elevated temperature that may be expected at the breakdown site \cite{antoine}.

The Black equation can be used to express the MTBF due to electromigration in small gaps in integrated circuits exposed to high gradients: 
\[ MTBF=B J^n e^{-E_a/kT}, \]
where $E_a, J, n, B$, and $kT$ are the electric field, the current density, an exponent, a constant, and the thermal dependence of the reaction, respectively.  We note that for field emission near 10 GV/m, $J \sim E^{14}$, so $R \sim E^{28}$ \cite{Black}.  

As mentioned elsewhere, mall
ohmic heating and thermal diffusivity, modified by the Nottingham effect (which also cools the surface) will determine the temperature of asperities exposed to high currents \cite{nottingham}.  Experiments with exploding wires have been done for many years, demonstrating that high current densities can easily vaporize small conductors; however, as the shape of the conductors departs from a cylinder, the thermal diffusivity must be considered.  As discussed above, we do not believe that Ohmic heating contributes significantly.

\section{Cavity Measurements}    

One of the experimental problems in studies of arcs is that most of the useful parameters vary over many orders of magnitude in a few nanoseconds.  The model presented in this paper shows, however, that all the parameters of the arc are, in principle, accessible to bothmodeling and experimental measurement, and a detailed comparison should be useful.

The power levels diverted into relativistic electrons can be measured from the x-ray radiation produced during breakdown events.  Since cavities are primarily thick-walled structures, the energy produced will generally be the result of electron  bremsstrahlung produced in the cavity walls. Since the radiation length of copper is 1.43 cm, comparable to the wall thickness, a significant fraction of a electron energy would be detected by radiation monitors located around the cavity.

We have measured the energy produced during breakdown events in an 805  MHz pillbox cavity roughly 0.08 m long at about 40 MV/m gradient.  At this field, the cavity contains about 1 J of electromagnetic energy.  Pits produced on the interior surface of the cavity have an average diameter of about 200 $\mu$m and depths of about 25 $\mu$m.  The total energy involved in melting and removing this material from the wall, neglecting the heat lost in the wall, was about 0.02 J, a small fraction of the available energy.  This energy is of the same order of magnitude as the plasma energy $U_p = nK(T_i+T_e)$ in the arc.

We have used a scintillator photomultiplier tube assembly with the voltage turned down until breakdown signals showed some height fluctuations, ensuring that the signals were not being saturated, to examine the rise time of the x-ray pulse produced by the shorting currents in the cavity. The results are shown  in Fig. 2b \cite{RCA}.  The photomultiplier detects x-ray radiation, which is presumed to be roughly proportional to the electron current accelerated across the cavity. Since this is  the largest source of energy loss during breakdown, we argue that this signal is proportional to the rate of energy loss of the cavity $P(t) \sim -dU/dt$.  We also plot the curve that was fitted to one of these breakdown waveforms.  If we integrate this signal, we can produce a signal proportional to the energy in the cavity, $U(t)$, whose derivative is proportional to the electric field or energy gained by electrons crossing the cavity, $V(t) \sim \sqrt{U(t)}$.  The shorting current is then $I(t)=-(dU/dt)/V(t)$.  We see that this current rises exponentially in the early stages of the discharge. The time constant of this  current rise should be related to the properties of the surface plasma; see Fig. 2c.  At least in the early stages, the plasma density is increasing with time.  Note that the overall time of a 1 J breakdown event is around 200 ns, so the instantaneous power is on the order 10 MW, assuming that the electric field $E$ is about 30 MV/m over a distance $d$ of 0.08 m, that the voltage $Ed$ is on the order of 2.5 MV, and that the current is $I=P/V \sim4$ A.  There is some scatter in the traces of photomultiplier signals,  so these numbers vary by perhaps a factor of 4 from event to event.

We fit the measured pulses with an expression (Fig. 2), 
\[ f(t) \sim \frac{1}{1+exp((t-t_1)/\tau_1))+exp((t_2-t)/\tau_2))}, \] 
where $t_1$ and $t_2$ determine the width of the pulse and $\tau_1$ and $\tau_2$ are the exponential  rise and fall times.  The exponential growth time can be determined from measurements of the leading edge of the photomultiplier pulse.  We assume that the rise time can be fitted from the initial rise in the x-ray pulse.  During reconditioning of the pillbox cavity, we recorded these pulse shapes and determined the time constants and the standard deviation of the distribution; these are plotted in Fig. 20.  The  figure shows that as the gradient (stored energy) increases, the growth time becomes shorter.   These measurements can be correlated with the estimates of growth times from modeling. However, the plasma growth rates through the discharge are governed by different mechanisms, initially by ionization times but, as the arc fully develops, by the available energy.

\begin{figure}  
\includegraphics[scale=0.5]{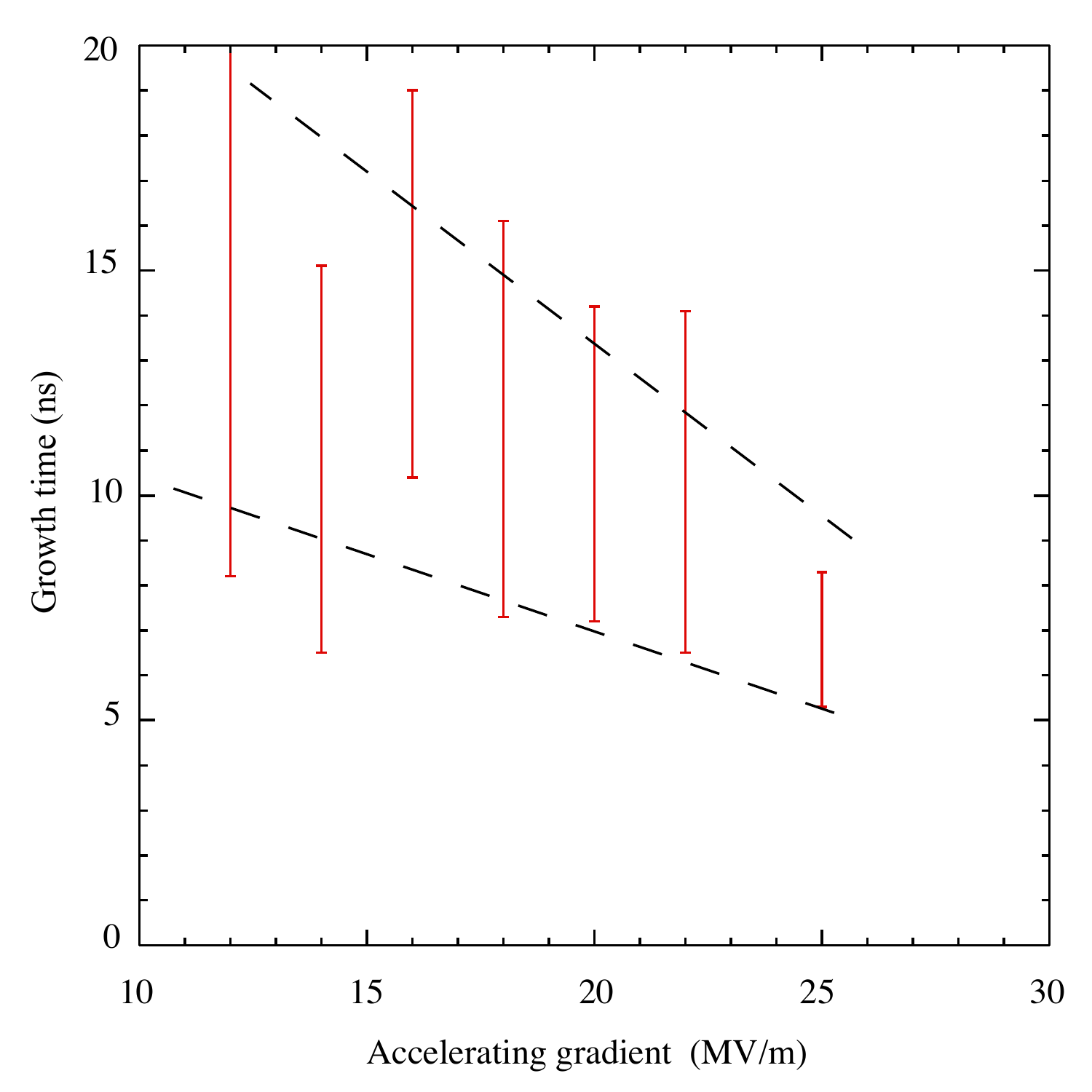}
\caption{Measured growth time of the x-ray pulse as a function of the accelerating gradient.  The dashed lines show the width of the distribution of rise times.}
\end{figure}

We can compare the data on growth times obtained from OOPIC Pro for the first few nanoseconds of the discharge and and the photomultiplier data at the end of the cycle.  Modeling gives an exponential growth time of about 1 ns for the initial few rf cycles.  The PMT data shows that the growth times are in the range of 3--30 ns, with the shortest times  produced only at the highest gradients.  To extend the range of the PMT data, we increased the high voltage on these tubes from 900 V to 1200 V to produce more gain and look at earlier in the pulse \cite{RCA}.  We obtained growth times on the order of 1.5 ns.  When these data are combined on the same plot, we obtain the results  shown in Fig. 21.  We find a continuous rise in the plasma density, with an initial time constant of about 1.5 ns.  The growth time constant slows slightly as the discharge gets more energetic, until it finally produces the shorting electron currents that produce the x-rays, when growth times of up to 50 ns were measured. 

\begin{figure}  
\includegraphics[scale=0.35]{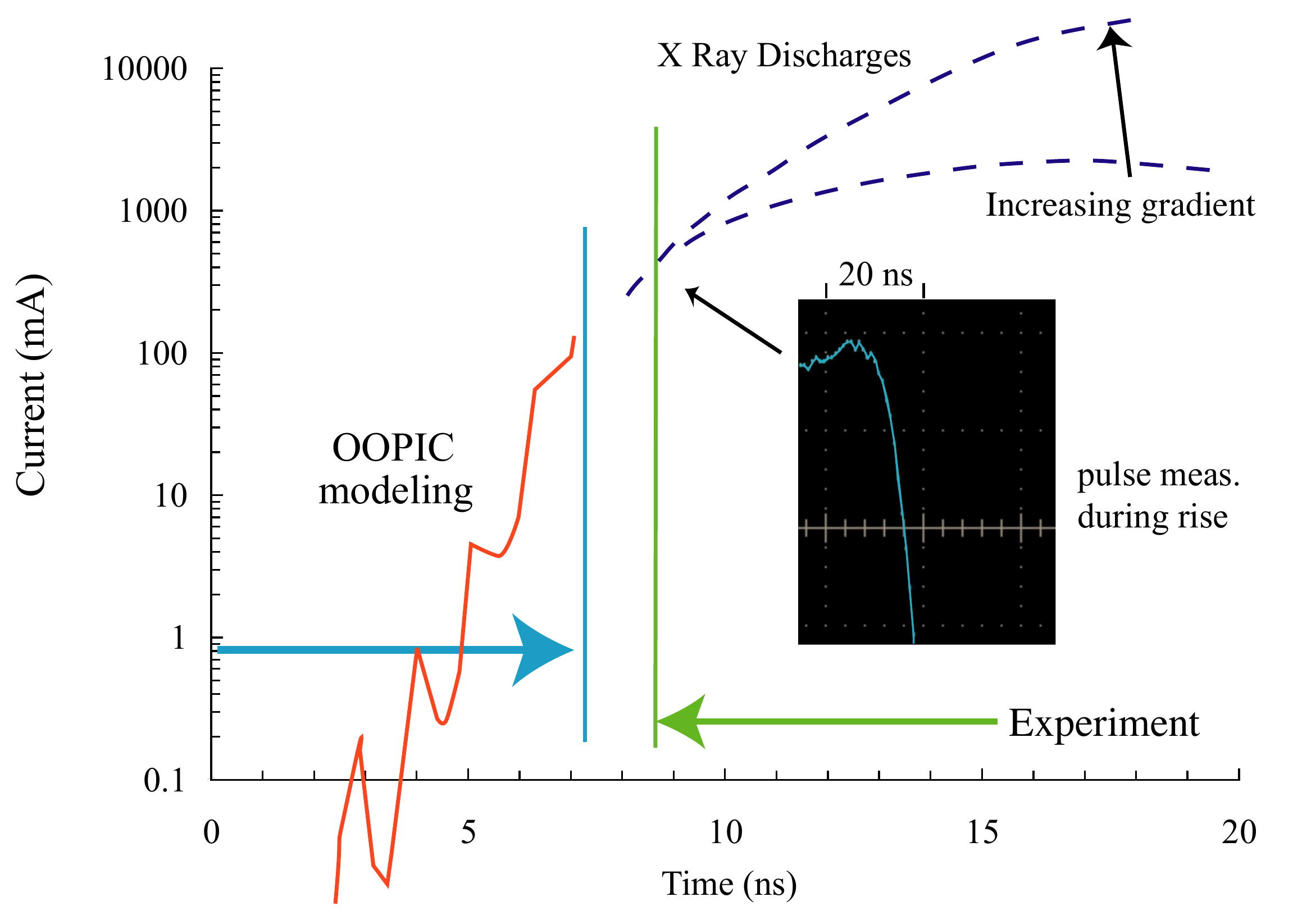}  
\caption{Simplified picture of the growth of the breakdown pulse.  OOPIC  Pro modeling
gives a $\sim$1 ns growth time in the initial  plasma charge density, and measurements of the x-ray pulse shape finally produced show a 3--30 ns growth time as the shorting pulse develops.  On the other hand, turning up the PMT gain (0.1 V/div.) shows growth times of 1.5--2.5 ns in the x-ray pulse before  reaching full current.  The vertical normalization of the OOPIC modeling is determined from measurements of electron motion. PMT data is normalized assuming a 1 J stored energy is discharged, giving A scale currents.  The inset shows the $\sim$1.5 ns risetime seen when the PMT gain was increased by a factor of $\sim$8 \cite{RCA}.}
\end{figure}

We believe an equilibrium develops between the surface damage created in breakdown events and the maximum field that can be obtained in a given cavity.  The overall influence of the surface damage left in the cavity by a breakdown event is described in Ref. \cite{hassanein}.  

One of the clear predictions of this model is that the only atoms involved in the formation of the plasma are copper and that any contaminants such as gasses or other metals would exist at a very low density in the plasma and in subsequent surface damage.  This prediction is confirmed by measurements of the spectroscopy of copper arcs.  These measurements cover the visible region of the spectrum and not the UV/x-ray region, where the the highest power densities of plasma radiation would be expected \cite{Dolgashev}.

\section{Discussion}

The field of vacuum arcs has been under study for over 110 years since the process was first identified.  Progress in understanding the mechanisms involved has been slow.  We believe that an attempt at a self-consistent approach, considering all stages of the discharge, has advantages both in identifying reasonable assumptions and in identifying mechanisms and ideas that seem directly or indirectly inconsistent with data.

The primary difference between the mechanism presented here and other models of arcing is that we assume the arc is triggered by electric tensile stresses, whereas others assume that the trigger is due to ohmic heating whiskers on the surface.  

The near-surface heating of the inside of an rf cavity by surface currents has been postulated as the primary trigger mechanism for rf breakdown \cite{Dolgashev}. However, this heating occurs primarily at the equator of a cavity, whereas breakdown occurs around irises, and the causal connection between these phenomena is not clear.

While the model we describe is somewhat similar to explosive electron emission (EEE) \cite{Mesiats}, the rf environment, with pulsed fields, is different from a DC arc, where the driving field is always present in the initial stages of the discharge.  It is not clear that the evolution of arcs in DC and rf fields should be significantly different.  A difference between this model and EEE, as described in Ref. \cite{Mesiats}, is the plasma pressure.  OOPIC Pro simulations give ion energies near the arc near or below 100 eV, and ion densities in the range of $10^{23}$ m$^{-3}$, giving a plasma pressure in the range of $10^6$ Pa, whereas the EEE model, which assumes that the plasma pressure fractures the surface, requires much higher plasma pressures in the emission zone to produce droplets. Our model of the breakdown arc assumes that the arc plasma that develops over the original asperity derives the majority of its power from field emission electrons that interact with the metallic plasma.  Field emission from hot materials has been described analytically by Jensen et al. \cite{jensen}. However, the primary limitation on the field emission current at all times is the space charge from a number of plasma species in motion in the immediate region above the field emitting surface, and this space charge limit is most easily evaluated numerically.   More detailed calculations are under way using OOPIC Pro and VORPAL \cite{OOPIC,VORPAL,OOPIC1}.

We believe this model is quite general and applies to a variety of initial states, including electron beam welding, micrometeorite impacts and arcs on tokamak rf antennas, where the initial state consists only of a dense plasma above a metallic surface \cite{PAC11}.

While we currently use PIC codes, we understand that these codes, which rely on two-body collisions, are inherently invalid at higher plasma densities and it is impossible to extrapolate to higher densities with these codes.  We are developing a numerical molecular dynamics/Monte Carlo theory of non-Debye plasma to apply to these rf problems.

\section{conclusions}  

This paper presents an evolving outline of a self-consistent set of mechanisms that seem to completely describe rf breakdown arcs.  We describe the process in terms of surface fracture, ionization of neutrals, plasma growth, and damage mechanisms.  We describe how high surface fields could fracture the  surface of field emitting asperities in an rf cavity to form a plasma. This plasma could provide sufficient electrons to melt the surface locally and short the cavity in a timescale of a few nanoseconds.  We believe the plasma modeling can, in principle, describe all aspects of the growth of the plasma and the mechanisms that drive and control this growth. We are exploring the possible range of predictions \cite{insepovz}.  

In an earlier paper we  described how an equilibrium between  the stored energy of the structure and the surface damage can determine many of the properties of an rf system \cite{hassanein}.  In this paper we have shown how the parameters  of the arc are determined by the rf environment.  For example, we show the amount of material expelled from the surface can determine whether an arc will occur. We have described a number of trigger mechanisms and concluded that electric tensile fracture (Coulomb explosions) as the dominant effect.  We have shown what mechanisms drive the development of the discharge; and we have shown how space charge, electron kinetics, and bulk heating can control the rate of the development of the discharge.  We have obtained good agreement between  predictions of our model and experimental measurements of the rise time of the x-ray pulse.
We have discussed why discharges can vary depending on the available energy.   

We have used OOPIC modeling to estimate very high ($>$1 GV/m) surface electric fields in the dense plasma, and we have measured fields of this order using electrohydrodynamic arguments to relate the dimensions of surface damage with the applied electric field.  We also have presented a geometrical picture of the large enhancement factors of field emitters that seems consistent with the absence of whiskers on surfaces exposed to high fields.  The enhancement factors we derived, when combined with the Fowler-Nordheim analysis, produce a consistent picture of breakdown and field emission from surfaces at local fields of 7--10 GV/m that are seen in other environments. We believe the general picture presented here for rf breakdown arcs should be directly applicable to a larger class of vacuum arcs. We see no evidence of significant ohmic heating in field emitters.

While this paper is primarily an outline of the mechanisms involved, it demonstrates that these mechanisms are capable of driving very fast avalanche processes that can interfere with or end the normal operation of the cavity. More detailed analysis of these mechanisms is under way, and we expect to be able to provide more precise results in the future.  Where possible we compare this model with  experimental data from various sources.  We believe all aspects of this model are experimentally accessible and a more detailed comparison of this model and experimental data would be very productive.

\section*{Acknowledgments}   

We  thank the staff of the Accelerator and Technical Divisions at Fermilab for supporting and maintaining The MAP experimental program in the MTA experimental area.    The work at Argonne is supported by the U.S. Department of Energy Office of High Energy Physics  under Contract No. DE-AC02-06CH11357.  The work of Tech-X personnel is funded by the Department of Energy under Small Business Innovation Research Contract No. DE-FG02-07ER84833.

\vspace{1 in}
\parbox{3.2 in}{The submitted manuscript has been created by UChicago Argonne, LLC, Operator of Argonne National Laboratory (``ArgonneÓ). Argonne, a U.S. Department of Energy Office of Science laboratory, is operated under Contract No. DE-AC02-06CH11357. The U.S. Government retains for itself, and others acting on its behalf, a paid-up nonexclusive, irrevocable worldwide license in said article to reproduce, prepare derivative works, distribute copies to the public, and perform publicly and display publicly, by or on behalf of the Government. }

\end{document}